\documentclass[fleqn,usenatbib]{mnras}

\usepackage{newtxtext,newtxmath}

\usepackage[T1]{fontenc}
\usepackage{ae,aecompl}

\usepackage{graphicx}	
\usepackage{amsmath}	
\usepackage{tabularx}
\usepackage{cleveref}
\usepackage{multirow, ulem}

\newcommand{\obj}{$\lambda$~And}

\newcommand{\msano}{~M_\odot~{\rm yr}^{-1}}

\title[Stellar wind of $\lambda$~And]{$\lambda$~And: A post-main sequence wind from a solar-mass star}

\author[D. \'{O} Fionnag\'{a}in et al.]{D. \'{O} Fionnag\'{a}in$^{1}$,
A.~A.~Vidotto$^{1}$\thanks{E-mail: aline.vidotto@tcd.ie},
P.~Petit$^{2}$,
C.~Neiner$^{3}$,
W.~Manchester~IV$^{4}$,
\and C.~P.~Folsom$^{2}$,
G. Hallinan$^5$
\\
$^{1}$School of Physics, Trinity College Dublin, College Green, Dublin 2, Ireland\\
$^{2}$IRAP, Université de Toulouse, CNRS, UPS, CNES, 14 Avenue Edouard Belin, 31400, Toulouse, France\\
$^{3}$LESIA, Paris Observatory, PSL University, CNRS, Sorbonne University, Universit\'{e} de Paris, 5 place Jules Janssen, 92195 Meudon, France\\
$^{4}$Department of Climate and Space Sciences and Engineering, University of Michigan, Ann Arbor, MI48109, USA\\
$^5$Department of Astronomy,  California Institute of Technology, Pasadena, CA 91125, USA
}

\date{Accepted XXX. Received YYY; in original form ZZZ}

\pubyear{2020}

\begin{document}
\label{firstpage}
\pagerange{\pageref{firstpage}--\pageref{lastpage}}
\maketitle

\begin{abstract}
We investigate the wind of $\lambda$ And, a solar-mass star that has evolved off the main sequence becoming a sub-giant. We present spectropolarimetric observations and use them to reconstruct the surface magnetic field of $\lambda$ And. Although much older than our Sun, this star exhibits a stronger (reaching up to 83 G) large-scale magnetic field, which is dominated by the poloidal component. To investigate the wind of $\lambda$ And, we use the derived magnetic map to simulate two stellar wind scenarios, namely a ``polytropic wind" (thermally-driven) and an ``Alfven-wave driven wind" with turbulent dissipation. From our 3D magnetohydrodynamics simulations, we calculate the wind thermal emission and compare it to previously published radio observations and more recent VLA observations, which we present here. These observations show a basal sub-mJy quiescent flux level at $\sim$5~GHz and, at epochs, a much larger flux density ($>37$~mJy), likely due to radio flares. By comparing our model results with the radio observations of $\lambda$ And, we can constrain its mass-loss rate $\dot{M}$. There are two possible conclusions. 1) Assuming the quiescent radio emission originates from the stellar wind, we conclude that $\lambda$ And has $\dot{M} \simeq 3 \times 10^{-9}$ M$_{\odot}$ yr $^{-1}$, which agrees with the evolving mass-loss rate trend for evolved solar-mass stars. 2) Alternatively, if the quiescent emission does not originate from the wind, our models can only place an upper limit on mass-loss rates, indicating that $\dot{M} \lesssim 3 \times 10^{-9}$ M$_{\odot}$ yr $^{-1}$. 
\end{abstract}

\section{Introduction}\label{sec:intro}
Stellar atmospheres are highly dynamic environments that change on timescales varying from milliseconds (e.g. flares) to giga-years (e.g. spin-down). {In a series of works \citep{OFionnagain2018, OFionnagain2018b, OFionnagain2019}, we have examined the evolution of winds of solar-type stars in the main-sequence phase. Here, we investigate the wind of a solar-type star after it has evolved off the main sequence. This star, $\lambda$~And, is a sub-giant   of spectral type G8 IV. It has a mass similar to that of our Sun, but a more inflated radius of $7.0~R_{\odot}$, a rotation rate of 54 days, and is at a distance of 24.2 pc (\Cref{tab_physical}). Being a solar-mass star, $\lambda$~And can help us contextualise the future evolution of the wind of our Sun. $\lambda$~And is a well studied star with X-ray \citep{Audard2003,Drake2011}, EUV \citep{Baliunas1984,Dupree1996,Sanz-Forcada2003}, optical \citep{Frasca2008}, interferometric imaging \citep{Parks2015}, and radio observations \citep[][see also Section \ref{sec.obs_radio} for more recent VLA observations]{Bath1976, Bowers1981, Lang1985}. This wealth of information makes $\lambda$~And a great candidate for our study, as we will use some of these observational results to better constrain the results of our wind simulations. }

{One key difference between \obj\ and the `future Sun' is that $\lambda$~And is likely more active than what the Sun will be at the post-main sequence. Once they evolved off the main sequence and their radii increase, conservation of angular momentum implies that (single) stars will spin down. Given that rotation and activity are related \citep{Skumanich1972, 2003A&A...397..147P, Vidotto2014, 2020MNRAS.495L..61L}, it is expected that these stars will become less active with age. Although the rotation of \obj\ is indeed slower than that of the Sun, its chromospheric Ca II H\&K activity is stronger \citep{Morris2019}. This increased activity is in line with its magnetic field, which, as we will show in this work, is stronger than the large-scale field observed in the present-day Sun. We will come back to this point further on this section.} 

Based on their coronal properties, cool stars that have evolved off the main sequence can be split into three distinct groups: sun-like stars with hot coronae, warm/weak coronal stars, and cold stars without coronae \citep{Linsky1979,1980ApJ...242..260H,Ayres2003,Cranmer2019}. Simply by placing $\lambda$~And on a HR-diagram we can see that, while somewhat evolved with a radius of 7.0 R$_{\odot}$, it has {probably} not yet lost its hot corona (\Cref{fig:cranmer_land}). Indeed, X-ray observations show that $\lambda$~And fits into the hot corona category as its spectrum shows hot line formations \citep{Linsky1979, Drake2011}. \citet{Ortolani1997} showed that the coronal temperature should exist around $0.9$ keV ($\approx 10.4$ MK), while \citet{Sanz-Forcada2003} found that during quiescence, the plasma temperature is closer to 7.9 MK. As a broad rule, stars in the `Hot Corona' region {(which includes the Sun) are believed to have mass-loss rates $\lesssim 10^{-10} \msano$  and terminal velocities $\gtrsim 400$ km s$^{-1}$. Stars to the right of this divide, with `No Corona', usually show larger mass-loss rates and terminal velocities of $\lesssim 40$ km s$^{-1}$ \citep{Drake1986,2016ApJ...829...74W,OGorman2018}. \Cref{fig:cranmer_land} shows a roughly smooth transition between these two groups. Stars that show signs of weak/warm coronae are part of an intermediate `Hybrid' group, perhaps giving rise to partially ionising winds, and having a combination of wind driving mechanisms. 

\begin{figure*}
    \centering
    \includegraphics[width=0.8\linewidth]{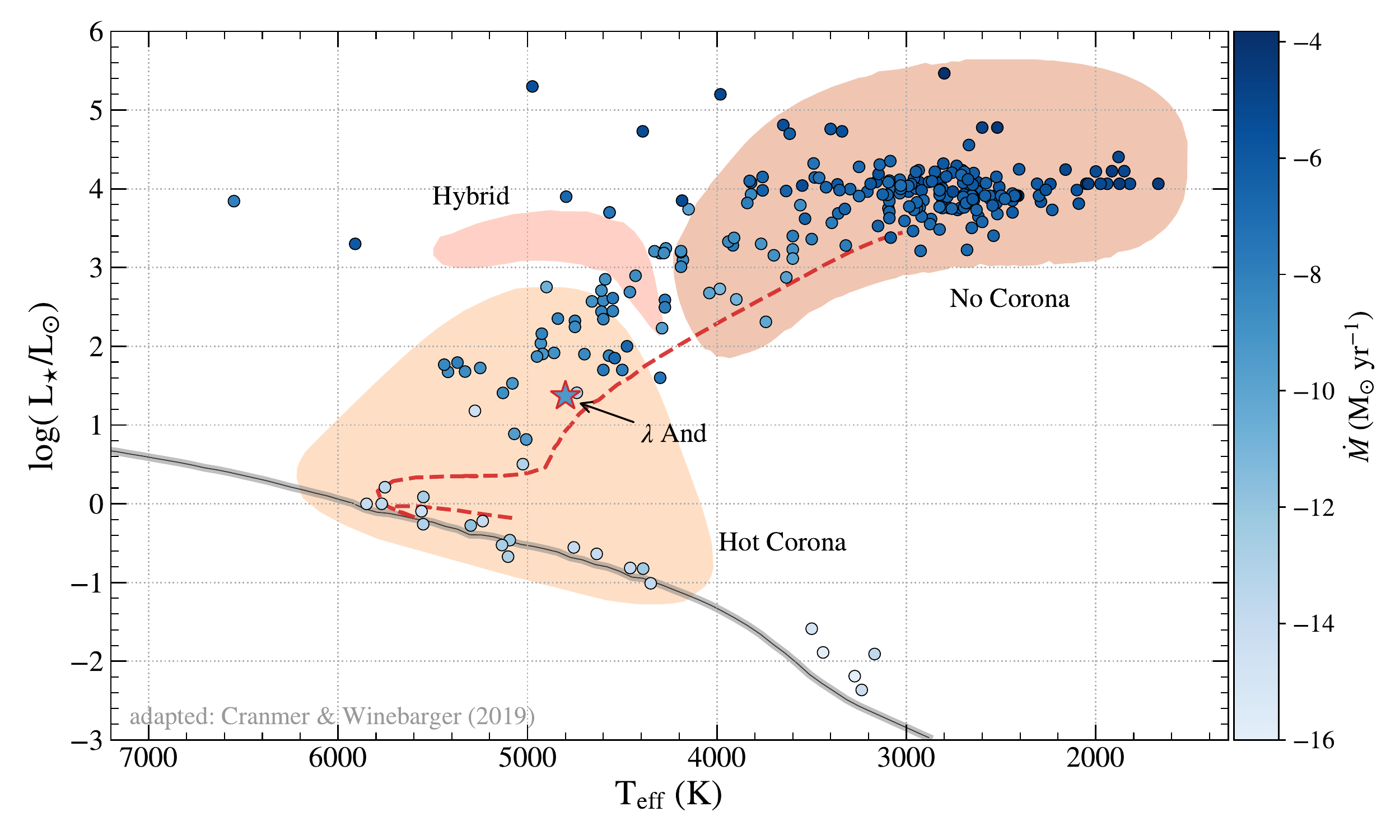}
    \caption{This is an adapted figure from \citet{Cranmer2019}. It shows the evolution of stellar mass-loss rates as low-mass stars evolve off the main sequence and become red giants. We contextualise the evolution of $\lambda$~And, shown over-plotted (red outlined star symbol). We see that although $\lambda$~And has begun to expand, it is still a sub-giant and retains its hot corona. Stellar mass-loss rate is shown as a blue scale. The grey line shows the zero-age main sequence, with filled regions shown for stars presenting hot corona, no corona, and a hybrid group \citep{Linsky1979,1980ApJ...242..260H,Ayres2003}. The dashed red line shows a 1-M$_{\odot}$ evolutionary track from \citet{Drake2011}.}
    \raggedright\footnotesize\textsuperscript{a} {Stellar data---priv. comm., S. Cranmer, Jan 2020.}
    \label{fig:cranmer_land}
\end{figure*}

\citet{Muller2001} and \citet{Wood2002} derived a mass-loss rate for the wind of $\lambda$~And indirectly through Ly-$\alpha$ absorption of excess neutral hydrogen build-up between the stellar wind and the astrosphere. The authors found a mass-loss rate $ 2 \times 10^{-13} \msano $ and $10^{-13} \msano $ respectively, although they claim the detection is uncertain. More recent work suggested an even lower mass-loss rate of $2 \times 10^{-15} \msano $ \citep{Wood2018}. This is unexpectedly low compared to the previously mentioned mass-loss rates for post-main sequence stars, and is a much lower mass-loss rate per unit surface area than the Sun itself. A particularly promising technique to constrain winds of low-mass stars  is to use radio observations of the thermal bremsstrahlung from their ionised winds \citep{Panagia1975, Wright1975, Lim1996, Villadsen2014, Fichtinger2017, Vidotto2017a, OFionnagain2019}. While stars along the main sequence possess winds too tenuous to detect with current instrumentation, the increased mass-loss rates of the more evolved low-mass stars provide a more attainable target (e.g. \citealt{OGorman2017}). {One of the goals of our work is to use radio observations of $\lambda$~And to constrain our wind models. Radio observations of $\lambda$~And were mainly published a few decades ago \citep{Bath1976,Bowers1981,Lang1985}. Here, we present more recent Karl G. Jansky Very Large Array (VLA) archival data for $\lambda$~And. We will use these radio flux densities to constrain our wind models in this work. }

\begin{table}
    \centering
    \caption{Physical parameters of $\lambda$~And from \citet{Drake2011}, {except for the rotational period, which is from \citet{1978AJ.....83..176L}.}}
    \begin{tabularx}{\columnwidth}{cccccc}
    \hline
    $M_\star$ & $R_\star$  & log(L$_{\star}$/L$_{\odot}$) & T$_{\rm eff}$  & $P_{\rm rot}$  & d  \\
     (M$_{\odot}$) & (R$_{\odot}$) & &  (K) &  (d) & (pc) \\ \hline \hline
    $1.0 \pm 0.2$ & $7.0 \pm 0.7$ & 1.37 & $4800$ & {$54.0\pm 0.5$} & $24.2\pm0.3$\\ \hline
    \end{tabularx}
    \label{tab_physical}
\end{table}

In addition to the wind models of \obj , we also present here the first full surface magnetic field observations of this star, finding a strong magnetic field for such an evolved star. These observations, carried out with the NARVAL spectropolarimeter, allow us to constrain the surface magnetic field of $\lambda$~And. These derived surface magnetic fields can constrain the lower boundary of the 3D magnetohydrodynamic wind simulations that we run. Usually, we see a decay in magnetic field strength as solar-type stars evolve, as their activity decreases along with their rotation \citep{Skumanich1972, Vidotto2014, Booth2020}. However, this sub-giant star seems to have a relatively strong large-scale magnetic field compared to the Sun. 
The exact process through which this star would reach this stage in its evolution with such a magnetic field {is} yet unknown. Potential reasons are that it began with a much stronger dynamo in its past than anticipated, or perhaps the secondary companion had some effect on the primary star at a point in the past. $\lambda$~And differs from the Sun as it is a RS Canum Venaticorum (RS CVn) variable, meaning it is a variable binary system. The variability on this star is likely due to magnetic spots coming in and out of view due to stellar rotation \citep{Baliunas1979, Baliunas1982, Donati1995, Henry1995, ONeal2001, Sanz-Forcada2001, Frasca2008, Drake2011}. {RS CVn systems, in particular, can present observed levels of chromospheric and coronal activity that are orders of magnitude higher than in single stars with similar spectral types \citep{1980ApJ...241..279A, 1981ApJ...245..671W}. This is likely caused by the increase in activity when the two stars interact with each other, which can  lead to rotational synchronisation of the system (e.g., \citealt{2004AN....325..393L}; an analogous process has been inferred to take place in close-in planet--star systems, \citealt{2000ApJ...533L.151C})}. For the purposes of this work, we assume the binarity of this system does not affect our wind models. Compared to the Sun, \obj\ is  metal-poor  ([Fe/H] $= -0.46 \pm 0.04$ dex, \citealt{Maldonado2016}). We do not include the effects of different metal abundances on the stellar wind and stellar evolution, but the effects of which have been examined in other works \citep{Suzuki2018}. 

In this work, we employ two 3D MHD wind models, using BATS-R-US \citep{Powell1999,Sokolov2013,VanDerHolst2014}, including the observed stellar magnetic field, with which we aim to more accurately replicate the observed radio flux of the wind of \obj . As stars age and progress along the red giant branch their outer atmospheres cool significantly, this cooling means they no longer have a hot corona to drive their stellar winds in the form of thermal acceleration. Despite this, the mass-loss rate of these stars dramatically increases by many orders of magnitude (see \Cref{fig:cranmer_land}). Therefore it is expected in these cool evolved scenarios, the wind is driven by waves, predominantly Alfven waves. It is also possible that winds have some sort of hybrid acceleration mechanisms, with characteristics from both a coronal driven hot wind, and cold wave-driven wind. In this work, {we use two wind models, namely a polytropic thermally-driven wind and an Alfven-wave driven wind, aiming at finding}  which one better reproduces the radio observations of \obj .

In \Cref{sec.obs}, we discuss the spectropolarimetric observations of the star{, which allow us to derive a surface magnetic field map using the Zeeman Doppler Imaging (ZDI) technique}. \Cref{sec:simulations} details the different models that we use to simulate the stellar wind of $\lambda$~And. {In \Cref{sec:discussion}, we present the results of our 3D MHD simulations. In Section \ref{sec:radio}, we use the observed radio emission of $\lambda$ And to select the most appropriate wind model. For that, we use both VLA archival observations (\Cref{sec.obs_radio}) and literature values.} We conclude and summarise our work  in \Cref{sec:conc}.

\section{Observed surface magnetic fields}  \label{sec.obs} 
$\lambda$~And was observed with the NARVAL high resolution spectropolarimeter installed on the Bernard Lyot Telescope (TBL, Pic du Midi Observatory, France, \citealt{Auriere2003}) in the frame of the BritePol program \citep{Neiner2017}. The circular polarisation mode of {NARVAL} was used to acquire the data, providing a simultaneous measurement of Stokes V and Stokes I over a wavelength domain extending from 370~nm to 1~$\mu$m at a spectral resolution of about 65,000. 

Each Stokes V sequence consists of 4 sub-exposures of 56 seconds each, obtained with different azimuthal angles of the half wave Fresnel rhombs in the polarimetric module \citep{Semel1993}. A null polarisation spectrum was also computed for each observation by destructively combining the 4 sub-exposures. This allows {a check} for any spurious signal in Stokes V that may have been produced by variable weather conditions, instrumental issues or non-magnetic stellar variations such as pulsations. 

The full set of BritePol observations consisted in six measurements obtained in December 2013, one in January 2014, and nineteen from August to October 2016. Our magnetic model was restricted to the 2016 data, as the long-term evolution of surface features on cool active stars similar to $\lambda$~And prevents us from combining data obtained over more than a few weeks (see e.g. \citealt{Petit2004a} for the active sub-giant primary of the RS CVn system HR 1099). We also removed from this time series the observation obtained on 10 August 2016, as the least squares deconvolution method (LSD, see paragraph below) led to an abnormal outcome for this specific spectrum. The subset selected here offers a good basis for tomographic mapping, with a dense set of observations spread over most of one stellar rotation (assuming a period of 54~d, \citealt{Drake2011}). All data used in this article are publicly available in the PolarBase data base \citep{Petit2014}. 

{Our Stokes V spectra do not exhibit clear signatures in any individual line, which is typical of the relatively small amplitude of Zeeman signatures recorded in most cool active stars, thus an approach combining many lines is needed. As usually done in this situation, we make use of the LSD method \citep{Donati1997, 2010A&A...524A...5K} }to extract an average, pseudo-line profile of enhanced signal-to-noise ratio. To do so, we adopt a list of lines produced by a photospheric model \citep{Kurucz1993} with stellar parameters close to those of $\lambda$~And ($T_{\rm eff} = 4800 \pm 100$~K and $\log g = 2.75 \pm 0.25$, \citealt{Drake2011}). We impose for the LSD pseudo-line profiles an equivalent wavelength of 650~nm, and an equivalent Land\'e factor of 1.21. The outcome is a time-series of Stokes I and Stokes V pseudo line profiles, with the systematic detection of a polarised signature at the radial velocity of the line (black points in \Cref{fig_stokesI,fig_stokesV}). 

The surface magnetic field geometry was calculated with the Zeeman-Doppler Imaging (ZDI) technique \citep{Semel1989}, using the spherical harmonics expansion proposed by \cite{Donati2006}, and the latest python implementation of \cite{Folsom2018d}. In this framework, the stellar surface is paved with rectangular pixels linked to a local line model. Following \cite{Folsom2018b}, the local Stokes I line profile takes the form of a Voigt profile weighted according to a projection factor and linear limb darkening coefficient (taken equal to 0.73, \citealt{Claret2004}). Each local line profile is also Doppler shifted according to the local radial velocity produced by stellar rotation, assuming $v \sin i = 7.3$~km~s$^{-1}$ \citep{Massarotti2008}. The local Stokes V line profile is computed from the local Stokes I profile and the local longitudinal field using the weak field approximation (where Stokes V is proportional to the first derivative of Stokes I). The global Stokes I and V profiles obtained after integrating over the visible stellar hemisphere are then Doppler shifted to follow the radial velocity variations produced by the orbital motion of the target. The width and depth of the local Voigt profiles are adjusted to match the observed set of Stokes I LSD pseudo-profiles. Our ZDI model includes spherical harmonics modes up to $\ell = 15$, as no noticeable improvement of the Stokes V fit is obtained when further increasing this number. The best ZDI model is obtained for an inclination angle equal to $71 \pm 2^\circ$, which is consistent (within uncertainties) with the estimate of \cite{Donati1995} ($60^{+30 \circ}_{-10}$). The sets of synthetic Stokes I and V profiles obtained with this ZDI procedure are illustrated in \Cref{fig_stokesI,fig_stokesV} as red solid lines.      

{Note that from the stellar radius of $7.0 \pm 0.7~R_\odot$ \citep{Drake2011},  rotation period of $54.0 \pm 0.5~$~d \citep{1978AJ.....83..176L}, and our derived inclination angle of $71 \pm 2^\circ$, we find a calculated $v \sin i = 6.20 \pm 0.63$~km~s$^{-1}$, which is about $1.8\sigma$ from the  value of  $v \sin i = 7.3$~km~s$^{-1}$ from \citet{Massarotti2008}. This apparent discrepancy disappears when one considers the uncertainty in the value from \citet{Massarotti2008}. Although these authors did not provide an uncertainty on their $v \sin i$, we estimate from the scatter in their Figure 5, that their uncertainty may be as large as 1~km~s$^{-1}$. Even if we take an error bar as small as 0.5~km~s$^{-1}$, the $v \sin i$ value derived from fundamental parameters ($R_\star$ and $P_{\rm rot}$) and our derived inclination is consistent with that from \citet{Massarotti2008} within error bars. }

The resulting magnetic geometry of \obj\ is plotted in \Cref{fig:zdimap}. Several magnetic spots are recovered and most of them are located near the equator (in both the radial and azimuthal field components). The maximum (local) field strength is equal to 83~G, while the average unsigned field strength is equal to 21~G. Most of the magnetic energy is reconstructed in the poloidal field component (64\%), and most of the poloidal field is observed in low order spherical harmonics components, with about 78\% of the poloidal magnetic energy in modes with $\ell \leq 3$. The low latitude azimuthal field forms a unipolar ring (of positive polarity), as already observed in several cool, evolved stars (e.g. \citealt{Donati2003,Petit2004a,Petit2004b}). We note that the rotation phases where the azimuthal field strength is large also have a strong radial field strength (around phase 0.2, and also between phases 0.6 and 0.9).

The reduced $\chi^2$ obtained by the ZDI inversion is equal to 1.9, showing that our simple magnetic model cannot fully reproduce the shape of the observed Stokes V pseudo profiles. We assume here a solid rotation of the stellar surface, and this assumption is often the main limitation of ZDI models of cool stars, since the surface is expected to be differentially rotating. Following the procedure of \cite{Petit2002}, we searched for a progressive distortion of the magnetic geometry under the influence of a solar or anti-solar surface shear. This search was inconclusive, likely because our data set does not cover more than one rotation period. Another possible limiting factor is the continuous appearance and disappearance of magnetic spots, and the relatively large timespan of our data (slightly less than 2 months) is possibly responsible for some significant intrinsic  evolution of the magnetic pattern.
\begin{figure}
    \centering
    \includegraphics[width=\linewidth]{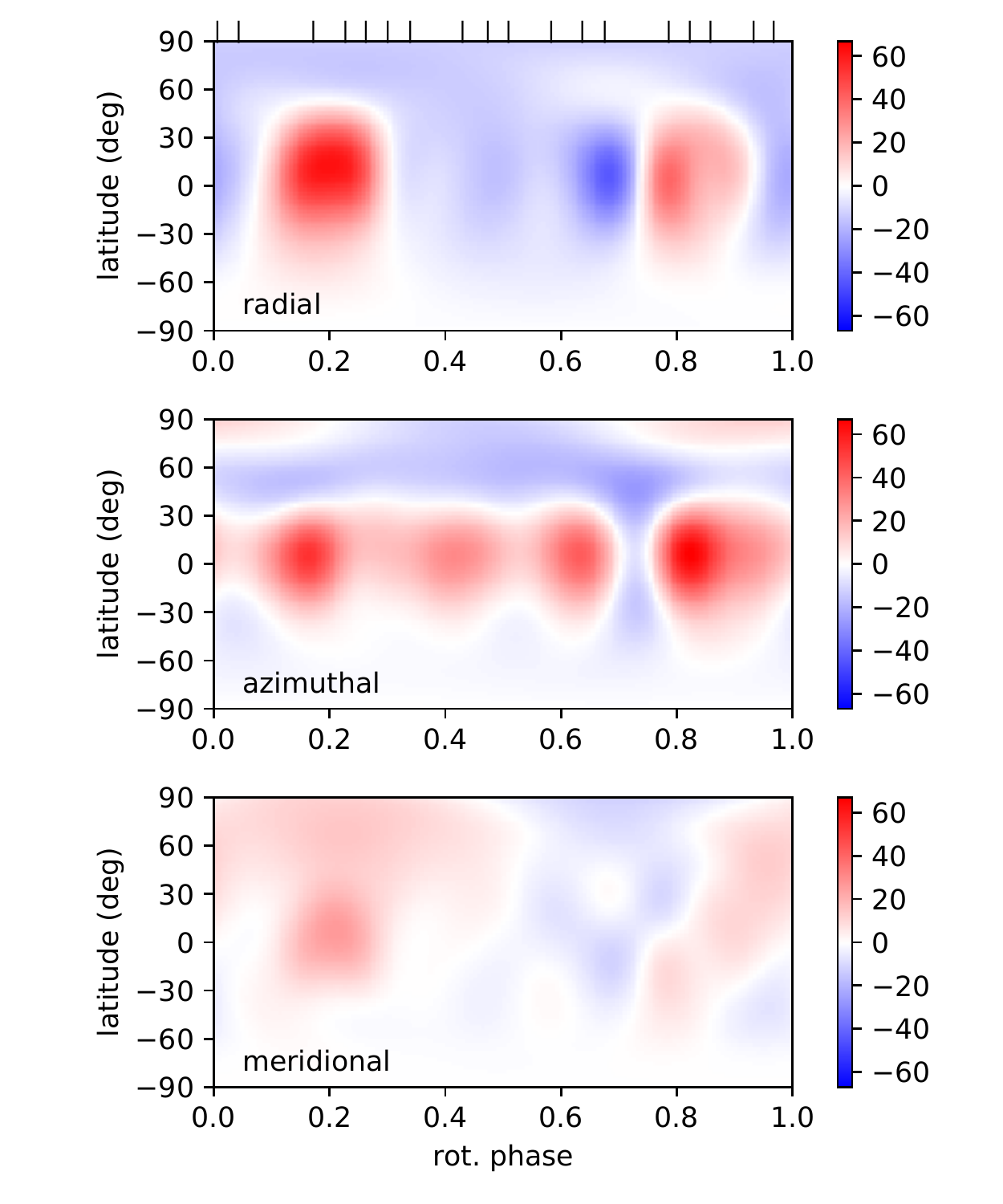}
    \caption{Large scale magnetic field geometry of $\lambda$ And, as reconstructed with the ZDI method. From top to bottom, the three panels show the radial, azimuthal and meridional components of the photospheric magnetic field (in Gauss). The observed rotational phases are shown as vertical ticks above the radial field map.}
    \label{fig:zdimap}
\end{figure}

To compare the surface magnetic field of $\lambda$~And with the magnetic survey for evolved stars of \cite{Auriere2015}, we have computed longitudinal field ($B_l$) values \citep{Rees1979} for every observation included in the ZDI analysis. The maximal longitudinal field strength throughout the time series is $|B_l|_{\rm max} = 13.7 \pm 0.4$~G, in good agreement with $|B_l|_{\rm max}$ values reported by \cite{Auriere2015} at similar rotation periods. We therefore suggest that, although $\lambda$~And is a member of a close binary system, the observed surface magnetic field strength is not noticeably influenced by the tidal interaction between the primary and its low-mass companion.    

{Magnetic field measurements of \obj\ have been obtained using the Zeeman broadening technique. \citet{Giampapa1983} reported an unsigned field strength of 1290~G covering 48\% of the surface, resulting in an unsigned average field of 619 G. \citet{1985ApJ...297..710G} found  600 G magnetic fields extending over at least 20\% of the visible hemisphere (an average of $\sim120$~G), but caution that their field strengths for \obj\ were not conclusive. In spite of the differences obtained in these two measurements, it is not surprising that field strengths measured with the Zeeman broadening technique are significantly higher than our unsigned average field strength derived by the ZDI technique of 21~G. This is because ZDI is limited to reconstructing the large-scale field, thus missing the small-scale field obtained in Zeeman broadening measurements \citep[e.g., ][]{2019ApJ...876..118S}.}

{Finally, photometric and spectroscopic observations have shown that $\lambda$~And is variable, with the presence of darker starspots being correlated with the brightening of Ca II K emission \citep{Baliunas1982}. This suggests that for more magnetically active periods there is a reduction in stellar brightness. Therefore, there may be a correlation between stellar magnetic geometry and V magnitude, which we did not explore here. An interesting future investigation would be to perform simultaneous photometric observations and ZDI mapping. }

\section{Models: Red giant stellar wind models}\label{sec:simulations}
We use two separate implementations of the BATS-R-US code, the {polytropic} wind model, as described in \citet{Vidotto2017a,OFionnagain2019} and the Alfven-wave driven  AWSoM model defined in \citet{VanDerHolst2014}. As evolved type stars possess cool extended atmospheres, we expect that they are wave driven (predominantly Alfven waves), which drive wind acceleration through turbulent dissipation. This concept has been used for evolved stars frequently in the past \citep{1980ApJ...242..260H, Vidotto2006, Suzuki2007, Airapetian2010, Cranmer2011, VanDerHolst2014, Yasuda2019}. For a star such as $\lambda$~And, it is possible that as it moves towards the hybrid area of \Cref{fig:cranmer_land}, the wind combines both thermal acceleration and wave driving. Therefore we carry out simulations of both cases to compare to observations. We summarise the essential equations to both models below:

\subsection{{Polytropic wind model (thermally driven)}}
\label{sec:polytropic}
In this model, the inner boundary of the simulation begins in the corona of the star. We assume a polytropic index which drives the wind of the star by supplying energy to the wind. The polytropic index in the solar wind has been measured as $\Gamma = 1.1$ \citep{VanDoorsselaere2011a}, and many numerical solar wind simulations use $1 < \Gamma < 1.15$ \citep{Keppens1999b,matt2012,Johnstone2015a,johnstone2015b}, here we adopt a value of $\Gamma = 1.05$. BATS-R-US solves for 8 fluid quantities in this case: mass density ($\rho$), wind velocity ($\textbf{u} = \lbrace u_x, u_y, u_z \rbrace$), magnetic field ($\textbf{B} = \lbrace B_x, B_y, B_z \rbrace $), and gas pressure P. The equations that govern this model are shown below.

\begin{equation}
 \frac{\partial \rho}{\partial t} + \nabla \cdot (\rho \textbf{u}) = 0,
 \end{equation}
 \begin{equation}
 \frac{\partial (\rho \textbf{u})}{\partial t} + \nabla \cdot \left[ \rho \textbf{uu} + \left( P + \frac{B^2}{8\pi} \right)I - \frac{\textbf{BB}}{4\pi} \right] = \rho \textbf{g},
 \end{equation}
 \begin{equation}
 \frac{\partial \textbf{B}}{\partial t} + \nabla \cdot (\textbf{uB} - \textbf{Bu}) = 0
 \end{equation}
 \begin{equation}
 \frac{\partial \varepsilon}{\partial t} + \nabla \cdot \left[ \textbf{u} \left(  \varepsilon + P + \frac{B^2}{8\pi} \right) -\frac{(\textbf{u} \cdot \textbf{B}) \textbf{B} }{4\pi} \right] = \rho \textbf{g} \cdot \textbf{u},
 \end{equation}
where the total energy density is given by:
\begin{equation}
 \varepsilon = \frac{\rho u^2}{2} + \frac{P}{\Gamma - 1} + \frac{B^2}{8\pi}
 \label{eq:state}
\end{equation}
Here, $I$ denotes the identity matrix, and \textbf{g} the gravitational acceleration. We assume that the plasma behaves as an ideal gas, that $P = n_{\rm cor} k_B T $, where $n_{\rm cor} = \rho / (\mu m_p)$ is the total number density of the wind. $\rho$ represents the mass density, k$_B$ is the Boltzmann constant, and $\mu m_p$ denotes the average particle mass. We take $\mu = 0.5$, which represents a fully ionised hydrogen wind. Polytropic index aside, the other free parameters are the base coronal density, base coronal temperature, and base magnetic field. For this model we use $n_{\rm cor} = 2.5 \times 10^{10} $ cm$^{-3}$ and $T_{\rm cor} = 1$ MK. The base magnetic field is constrained using ZDI observations, of which we only include the radial field component, B$_r$, in our simulations, which are thoroughly described in \Cref{sec.obs}. {The singular {polytropic} wind scenario listed in \Cref{tab:AWSoM} (\textbf{Z0}) has a 1 MK base temperature as the simulation begins embedded in the corona.} Note that all parameters here equate to coronal values, as this is where the bottom of this simulation begins. As we will show in Section \ref{sec.result_poly}, the reason why we chose this particular base density is because it  reproduces values of quiescent radio flux densities of \obj . We use a Cartesian grid, with the minimum resolution of 0.01 $R_{\star}$ and a maximum resolution of 0.3 $R_{\star}$, totalling 622,672 blocks, or $3.98 \times 10^{7}$ cells.

\subsection{Alfven wave-driven wind model with turbulent dissipation}
We use the SC (solar corona) module of the AWSoM code to simulate the Alfven-wave driven wind scenario. This module of the SWMF framework has been used previously to simulate the Alfven-wave driven wind of the Sun \citep{Sokolov2013,VanDerHolst2014, Meng2015,Oran2017,Gombosi2018} and other main sequence stars \citep{Alvarado-Gomez2018,Alvarado-Gomez2019, 2020A&A...635A.178B}. In this model an Alfven wave flux is assumed to be propagating from the base of the wind. Wave dissipation follows from a turbulent cascade resulting from the interaction of forward propagating and reflected waves. The equations that differ from the polytropic model described in \Cref{sec:polytropic} are the momentum equation, which includes separated electron ($P_e$) and ion pressures ($P_i$), and the additional pressure from the Alfven waves ($P_A$)
\begin{equation}
 \frac{\partial (\rho \textbf{u})}{\partial t} + \nabla \cdot \left[ \rho \textbf{uu} + \left( P_i + P_e + P_A + \frac{B^2}{8\pi} \right)I - \frac{\textbf{BB}}{4\pi} \right] = \rho \textbf{g}.
\end{equation}
The energy equations for electrons and ions become, respectively 
\begin{align}
 \frac{\partial \varepsilon_i}{\partial t} &+ \nabla \cdot \left[ \textbf{u} \left(  \varepsilon_i + P_i + \frac{B^2}{8\pi} \right) -\frac{(\textbf{u} \cdot \textbf{B}) \textbf{B} }{4\pi} \right] \notag \\ &= -(\textbf{u} \cdot \nabla)(P_e + P_A) + \frac{n_i k_B}{\tau_{ei}} (T_e - T_i) + Q_i + \rho \textbf{g} \cdot \textbf{u}
\end{align}
\begin{align}
    \frac{\partial}{\partial t} \left( \frac{P_e}{\gamma -1} \right) &+ \nabla \cdot \left( \frac{P_e}{\gamma - 1} \textbf{u} \right) + P_e \nabla \cdot \textbf{u} \notag \\ &= -\nabla \cdot \mathbf{q}_e + \frac{n_i k_B}{\tau_ei}(T_i - T_e) - Q_{\rm rad} + Q_e
\end{align}
where $\varepsilon_i$ represents the energy for the ions, according to \Cref{eq:state}. T$_{e,i}$ and n$_{e,i}$ denote electron and ion temperatures and number densities, respectively. We employ the equation of state $P_{e,i}$ = n$_{e,i}$ k$_B$ T$_{e,i}$ and the adiabatic index is $\gamma$ = 5/3. $\mathbf{q}_e$ represents the electron heat transport which transitions smoothly from collisional \citep{Spitzer1953} to collisionless \citep{Hollweg1978} heat flux so that the Spitzer-Harm collisional form dominates near the star, and the Hollweg collisionless form dominates further out in the wind. $Q_e$ and $Q_i$ are the heating functions for electrons and ions, respectively, and are partitioned forms of turbulent dissipation by Alfven waves \citep{Chandran2011}. $Q_{\rm rad}$ is the radiative cooling 
\begin{equation}
    Q_{\rm rad} = \Lambda n_e n_i, 
    \label{eq:cooling}
\end{equation}
where $\Lambda$ is the radiative cooling rate from CHIANTI v9.0 \citep{Dere2019}. 

{We follow the same prescription used for the solar wind simulations, in which we broaden the transition region by a factor $f=(T_m/T_e)^{5/2}$, with $T_m = 2.2 \times 10^5$~K  \citep{Sokolov2013,VanDerHolst2014}. With this transition region model, the energetic processes of heat conduction, radiative cooling and wave dissipation are modified by a factor $f$ everywhere where $T_e<T_m$: the heat conduction coefficient is increased by a factor $f$, while the radiative cooling and wave dissipation length scale are decreased by a factor $f$  (Equation 41 in \citealt{Sokolov2013}). These transformations do not change the temperature profile, but the result of them is an artificial increase in the extension of the transition region by a factor $f$ \citep{Sokolov2013}, which can be more easily modelled numerically. }

The Alfven wave dissipation, reflection and propagation are governed by the wave energy density equations 
\begin{equation}
    \frac{\partial w_{ \pm}}{\partial t}+\nabla \cdot\left[\left(\mathbf{u} \pm \mathbf{V}_{A}\right) w_{ \pm}\right]+\frac{w_{ \pm}}{2}(\nabla \cdot \mathbf{u})=\mp \mathcal{R} \sqrt{w_{-} w_{+}}-\xi_{ \pm} w_{ \pm}
\end{equation}
where $w_{ \pm}$ represents the wave energy densities for waves parallel ($+$) and anti-parallel ($-$) to the magnetic field. $\textbf{V}_A =\mathbf{B} / \sqrt{4\pi \rho}$ is the Alfven velocity, $\mathcal{R}$ is the wave reflection rate, and $\xi$ is the dissipation rate, given by
\begin{equation}
    \xi_{ \pm}=\frac{2}{L_{\perp}} \sqrt{\frac{w_{\mp}}{\rho}}
    \label{eq:dissipation}
\end{equation}
where $L_{\perp}$ is the transverse correlation length of the Alfven waves perpendicular to \textbf{B}. As in \citet{Hollweg1986}, $L_{\perp} \propto B^{-1/2}$, with the proportionality constant ($\ell$) set as an adjustable parameter of the model. The reflection rate $\mathcal{R}$ depends on the ratio of energy densities of parallel and anti-parallel waves, and the Alfven velocity.  {The inner boundary condition for the  wave energy density is $w = S_A/V_A = (S_A/B) \sqrt{4\pi \rho}$, where  $S_A$ is  the Poynting flux of the waves, with all values imposed at the inner boundary (surface of the star). The adjustable parameter of the model is the flux-to-field ratio $(S_A/B)$.} A thorough description of this entire model can be found in \citet{VanDerHolst2014}. 

The model requires values to be set for the free parameters, which range from the chromospheric density ($n_{\rm chr}$), chromospheric temperature ($T_{\rm chr}$), the Poynting flux-to-field ratio ($S_A/B$), and the damping proportionality constant ($\ell = L_{\perp} B^{1/2}$). {We discuss below how each of these are chosen in our models.}

 {There have been some works that constrained the density and temperature in the chromosphere of $\lambda$~And. For example, \citet{Sanz-Forcada2001} estimated plasma densities of $2 \times 10^{12}$~cm$^{-3}$ for $\lambda$~And, which is in the middle (in log scale) of the range of chromospheric densities we select {(from $1.5\times10^{10}$~cm$^{-3}$  to $1.5 \times 10^{14}$~cm$^{-3}$)}. \citet{Baliunas1979b} suggested chromospheric temperatures > 10,000 K and high coronal temperatures were observed in the EUV by \citet{Sanz-Forcada2001}. This is similar to what we observe in the solar atmosphere.  We thus use a typical solar chromospheric temperature of 50,000 K in our models. Some of the free parameters in our model can be limited \textit{a posteriori}. For example, selecting a base density that is too large could cause an unrealistically high mass-loss rate and the estimated radio emission could exceed observed levels (see \Cref{sec:radio}). 
 
The  physical parameters for the waves, namely $S_A/B$ and $\ell = L_{\perp} B^{1/2}$, are certainly less constrainable from observations. For a low-gravity star with $\log g=3$ (similar to \obj ), and an effective temperature of 4800K, the models by \citet{2002A&A...386..606M}  estimate a wave flux on the order of $10^8$~erg~cm$^{-2}$~s$^{-1}$ (we quote the results shown in their figure 8, which adopts their standard parameters, with mixing length $\alpha=2$ and magnetic field that is 85\% of the equipartition field; note however that lower fluxes are expected for $\alpha<2$). Other works on Alfven-wave driven wind models of giant stars have used wave fluxes on the order of $10^6 - 10^7$~erg~cm$^{-2}$~s$^{-1}$ \citep{1980ApJ...242..260H, Suzuki2007}, which is roughly of the same order of magnitude as those adopted in some works for red supergiants \citep{1989A&A...209..327J, Vidotto2006, Airapetian2010}. Guided by these studies, in our simulations, we chose Poynting flux-to-field ratio\footnote{We use a Poynting flux-to-field ratio that ranges between $S_A/B = 3.7 \times 10^4$ to $1.1 \times 10^{7}$ W~m$^{-2}$~T$^{-1}$. We use SI units as it is easier to compare with previous works which employed AWSoM. For example, solar wind  simulations by \citet{VanDerHolst2014} and \citet{Oran2017} have  adopted values of  $1.1 \times 10^{6}$ and $7.6\times 10^{5}$~W~m$^{-2}$~T$^{-1}$, respectively.  The radial magnetic field of \obj\ has a maximum field strength of $\sim 60$~G = $6\times 10^{-3}$~T, which  means that the maximum Pointing flux at the surface of the star ranges between $S_A\sim 2.2\times 10^2$   and    $6.7 \times 10^4$  W~m$^{-2}$. However, to compare with literature on Alfven-wave driven winds, it is more convenient to express this input parameter in terms of the Alfven wave flux in cgs units, resulting in maximum surface wave energy fluxes ranging between $2.2\times 10^5$ to  $6.7\times 10^7$~erg~cm$^{-2}$~s$^{-1}$.} $S_A/B = [3.7 \times 10^4,1.1 \times 10^{7} ]$~W~m$^{-2}$~T$^{-1}$, which translates to wave energy flux $S_A \simeq [2.2\times 10^5, 6.7\times 10^7]$~erg~cm$^{-2}$~s$^{-1}$. Note that our chosen range also includes previously used values in solar wind simulations with AWSoM \citep{VanDerHolst2014, Oran2017}.

The other input parameter adopted in our model is the scaling related to the correlation length $\ell = L_{\perp} B^{1/2}$. \citet{Hollweg1986} discussed that this value can be related to the distance between magnetic flux tubes on the stellar surface. They estimated a value of $\ell = 7520~{\rm km}\sqrt{\rm G}$ for the Sun empirically. Simulations of the solar wind previously done using AWSoM have adopted values for the Sun in the range $ \ell = [0.25, 1.5] \times 10^5~{\rm m}\sqrt{\rm T} = [0.25, 1.5] \times 10^4~{\rm km}\sqrt{\rm G} $ \citep{Sokolov2013, VanDerHolst2014, Oran2017}. We do not know how the parameter $\ell$ would change for \obj . If we use the physical reasoning from  \citet{Hollweg1986}, in which this number could be related to the distance between magnetic flux tubes on the stellar surface, we naively expect that for \obj , which is a star that has a radius that is seven times larger than that of the Sun, $\ell$ would be larger than the value adopted for the solar wind. Therefore, we use three different values of this parameter in our models: the value of $\ell_0 = 1.5 \times 10^5~{\rm m}\sqrt{\rm T}$ adopted for the solar wind \citep{VanDerHolst2014}, $7 \, \ell_0$, and a much larger value of $253 \, \ell_0$.
 
For our {Alfven-wave driven wind}, we run a number of simulations varying these input parameters, as shown in \Cref{tab:AWSoM}. We begin these simulations with a maximum dipolar magnetic field of $60$ G, which is similar to the maximum field strength in the radial component of the ZDI map (\Cref{fig:zdimap}). Additionally, we run a set of simulations using the ZDI map for the {Alfven-wave driven wind}, three of which are shown in \Cref{tab:AWSoM}: \textbf{C1}, \textbf{C2}, and \textbf{C3}. 

Our polytropic wind simulations reach steady state after a few tens of thousand iterations. However, this does not happen in some of the Alfven-wave driven wind simulations, which reach a quasi-steady state instead. This occurs as the heating depends on the dissipation of Alfven waves, which in turn depends on the magnetic field geometry and strength, the simulations tend to reach a point where they oscillate.  For example, if we take case \textbf{C1}, the maximum radial wind velocity in the $xz$-plane varies from 572 to 593~km/s, within an interval of 26000 iterations, while for case \textbf{D1}, the simulations reach a more steady solution (with variations in velocity of only a few km/s).  In the cases where a steady-state solution was not found, an average of the states is taken for the simulation parameters shown in Table \ref{tab:AWSoM}.

The SC module in the Alfvén wave-driven model uses a 3D spherical grid, with radial stretching from 1 to 30 $R_{\star}$. It also employs adaptive mesh refinement (AMR), adding extra refinement to the volume surrounding the current sheet. Radial stretching and AMR are quite efficient, increasing the resolution near the star and in required locations, without significantly increasing the number of cells in the simulation. The AMR is turned on for a single timestep after 100 timesteps to add refinement to the current sheet, which is the region where the magnetic field changes polarity and is susceptible to magnetic reconnection and high currents, which could cause issues in simulations without AMR. Our simulation mesh has r$_{\rm min,max} = 0.0003, 1.25 R_{\star}$ and $\phi/\theta_{\rm min,max} = 0.025, 1.5 R_{\star}$, resulting in an average of 45k blocks, and $4.3 \times 10^{6}$ cells.

\renewcommand{\arraystretch}{1.2}
\begin{table*}
\centering
\caption{Summary of our simulations. Cases \textbf{A},  \textbf{B},  \textbf{C} and  \textbf{D} refer to the Alfven-wave driven model, while \textbf{Z} refers to the polytropic wind model.  For the  Alfven-wave driven simulations, the base is set at the chromosphere, with a base temperature of 50,000 K, while the polytropic model has a base starting at the corona. For easy comparison with previous AWSoM simulations, we present the wave-related inputs in units used in solar wind simulations of \citet{VanDerHolst2014}: $\mathcal{S}_0=1.1\times 10^6$ W m$^{-2}$T$^{-1}$  and $\ell_0 = 1.5 \times 10^5$ m T$^{1/2}$, as well as their values in cgs units. Radio flux densities $\Phi_{\rm radio}$ were computed at 4.5~GHz. {The simulations with a dipolar topology has a polar field strength of 60~G.} }
\begin{tabular}{ccccccc|ccccc}
\multicolumn{7}{c}{Simulation Input} & \multicolumn{5}{c}{Simulation Output} \\
\hline
 ID & B field & $n_{\rm base}$ & $S_A/B$ & $S_A/B$ & $\ell$  & $\ell$  & $v_{\rm max}$ & $T_{\rm max}$ & $\dot{M}$ & $\dot{J}$ & $\Phi_{\rm radio}$ \\ 
  Sim & geometry & [cm$^{-3}$] & [$\mathcal{S}_0$] &  [$10^4$ erg cm$^{-2}$G$^{-1}$] & [$\ell_0$] & [$10^4~\textrm{km}\sqrt{\textrm G}$]  & [km s$^{-1}$] &  [MK] &  [M$_{\odot}$ yr$^{-1}$] &  [erg] &  [mJy]\\ 
  \hline \hline
 \textbf{A0} & Dipole & \multirow{4}{*}{$1.5 \times 10^{10}$} & 0.34 & 3.7 & 1 & 1.5& 700 & 5.8 & $4.7 \times 10^{-12}$ & $3.1 \times 10^{34}$ & 0.005\\
 \textbf{A1} & Dipole &                                                          & 3.4 & 37& 1 &1.5& 930 & 11 & $3.9 \times 10^{-11}$ & $1.1 \times 10^{35}$ & 0.011\\
 \textbf{A2} & Dipole &                                                          & 0.34 & 3.7& 7 &10.5& 817 & 6.6 & $4.1 \times 10^{-12}$ & $3.0 \times 10^{34}$ & 0.004\\
 \textbf{A3} & Dipole &                                                          & 0.034 & 0.37& 7  &10.5& 405 & 3.37 & $5.6 \times 10^{-13}$ & $1.6 \times 10^{34}$ & 0.004\\ \hline
 \textbf{B0} & Dipole & \multirow{2}{*}{$1.5 \times 10^{12}$} & 0.34 & 3.7& 7  &10.5& 765 & 5.7 & $3.7 \times 10^{-12}$ & $3.2 \times 10^{34}$ & 0.007\\
 \textbf{B1} & Dipole &                                                          & 0.34 & 3.7& 253  &380& 977 & 4.9 & $2.0 \times 10^{-12}$ &  $2.5 \times 10^{34}$ & 0.008\\ \hline
 \textbf{C0} & Dipole & \multirow{4}{*}{$1.5 \times 10^{13}$} & 0.34 & 3.7& 253 &380& 497 & 4.4 & $1.9 \times 10^{-11}$ & $3.7 \times 10^{34}$ & 0.014\\
 \textbf{C1} & ZDI &                                                              & 0.34 & 3.7& 253  &380& 590 & 2.9 & $1.2 \times 10^{-11}$ & $9.8 \times 10^{33}$ & 0.014\\
 \textbf{C2} & ZDI &                                                             & 3.4 & 111& 7  &10.5& 619 & 4.3 & $1.7 \times 10^{-10}$ & $4.5 \times 10^{34}$ & 0.030\\ 
 \textbf{C3} & ZDI &                                                             & 0.034& 0.37& 253  &380& 527 & 2.1 & $1.7 \times 10^{-12}$ & $3.0 \times 10^{33}$ & 0.007\\ \hline
 \textbf{D0} & Dipole & \multirow{2}{*}{$1.5 \times 10^{14}$} & 0.34 & 3.7& 253  &380& 765 & 5.6 & $4.0 \times 10^{-12}$ & $3.1 \times 10^{34}$ & 0.014 \\ 
 \textbf{D1} & Dipole &                                                           & 0.034 & 0.37& 253 &380& 292 & 1.7 & $2.7 \times 10^{-12}$ & $1.8 \times 10^{34}$ & 0.013\\ \hline \hline 
 \textbf{Z0} & ZDI & $ 2.5 \times 10^{10}$ & --- & ---& --- & --- & 456 & 1.0 & 2.9 $\times 10^{-9}$ & $2.2 \times 10^{35}$ & 0.890~\\\hline
\label{tab:AWSoM}
\end{tabular}
\end{table*}

\subsection{Mass and angular momentum losses}
From our wind simulations we can calculate the mass-loss rate from each of the stars by integrating the mass flux through a spherical surface $S$ around the star
\begin{equation}
    \dot{M} = \oint_S \rho u_r dS, 
    \label{eq:mdot}
\end{equation}
where $\dot{M}$ is the mass loss rate, $\rho$ is the wind density, and $u_r$ is the radial velocity. We also determine angular momentum-loss rate from our simulations as
\begin{equation}
    \dot{J} = \oint_S \left[ -\frac{\varpi B_{\phi} B_r}{4\pi} + \varpi u_{\phi} \rho u_r \right] dS
\end{equation}
where B and u are the magnetic field and velocity components of the wind, $\varpi = (x^2+y^2)^{1/2}$, the cylindrical radius, and r and $\phi$ denote the radial and azimuthal components respectively \citep{Mestel1999, Vidotto2014b}. Both mass-loss and angular momentum-loss are calculated from our simulations. 

\subsection{Radio modelling}\label{sec:radioequations}
Stellar winds emit thermal bremsstrahlung as they are composed of ionised plasma. Initially, the calculated radio flux from these winds was done analytically \citep{Panagia1975,Wright1975,Lim1996}, but with 3D simulations, it has become possible to do this calculation numerically \citep{Moschou2018,Cohen2018,OFionnagain2019}. From our simulations we can calculate the expected radio flux density by analysing the particle density and temperature structure in the wind. We use the Python package {\footnotesize {RadioWinds}} \citep{OFionnagain2018b} to calculate the thermal radio emission from the wind of $\lambda$~And. We can calculate the thermal emission expected from these winds by solving the radiative transfer equation, 
\begin{equation}
    I_{\nu} = \int_{0}^{\tau'_{\rm max}} B_{\nu} e^{-\tau_{\nu}} d\tau'
    \label{eq:intensity}
\end{equation}
where I$_{\nu}$ denotes the intensity from the wind, B$_{\nu}$ represents the source function, which in the thermal case becomes a blackbody function, $\tau_{\nu}$ represents the optical depth of the wind, with $\tau'$ representing our integration coordinate across the grid. The optical depth of the wind depends on the absorption coefficient, $\alpha_{\nu}$, of the wind as 
\begin{equation}
	\tau_{\nu} = \int \alpha_{\nu} ds, 
	\label{eq:tau}
\end{equation}
where \textit{s} represents the physical coordinate along the line of sight, $\alpha_{\nu}$ is described as \citep{Panagia1975,Wright1975,Cox2000}, 
\begin{equation}
    \alpha_{\nu} = 3.692\times10^8 [1 - e^{-h\nu / k_BT}] Z^2 f_g T^{-0.5}\nu^{-3} n_e n_i
    \label{eq:absorption}
\end{equation}
and the blackbody function is the standard Planck function.
\begin{equation}
    B_{\nu} = \frac{2h\nu^3}{c^2} \frac{1}{e^{hv/k_B T} -1}
    \label{eq:blackbody}
\end{equation}
where $\nu$ is the observing frequency, T is the temperature of the wind, h is Planck's constant, f$_g$ is the gaunt factor, Z is the ionic state of the wind, and n$_e$ and n$_i$ represent the electron and ion number densities of the wind \citep{Cox2000}. From the intensity we can calculate the flux density (S$_{\nu}$) of the wind as, 
\begin{equation}
    S_{\nu} = \frac{1}{d^2} \int I_{\nu} d\Omega
\end{equation}
where d is the distance to the star, and $\Omega$ is the solid angle.

\section{Results: Post-MS winds derived from our 3D MHD simulations}\label{sec:discussion}
{The last five columns in Table \ref{tab:AWSoM} show some of the outputs of our simulations. We take two separate approaches in this work to simulate the wind of $\lambda$~And, using a polytropic wind model (Z0) and an Alfven-wave driven wind model (all other cases in Table  \ref{tab:AWSoM}). In this section, we go through the main physical characteristics derived in each set of simulations. The key point of our paper, on the comparison between our calculated radio emission and the observed ones, will be presented in Section \ref{sec:radio}. } 

\subsection{Polytropic wind simulations}\label{sec.result_poly}
The simulation temperature and wind velocity for the polytropic wind model Z0 are shown in Figure \ref{fig:hotwind_tecplot}. This polytropic simulation displays a mass-loss rate of $2.9 \times 10^{-9}$ $\dot{M}_{\odot}$ yr$^{-1}$ or $1.3\times 10^5$ times the solar mass-loss rate. This value is much greater than astrospheric estimates ($2 \times 10^{-15}$ -- $10^{-13} \msano $, \citealt{Muller2001,Wood2018}), even though our terminal wind velocities ($\approx 400$ km s$^{-1}$) agree with the one adopted by these authors. Our mass-loss rate is more in line with the mass-loss rates of neighbouring stars in the HR diagram (\Cref{fig:cranmer_land}). Interestingly, our calculation for $\dot{M}$ would be in line with the X-ray---$\dot{M}$ trend for pre-wind dividing lines in \citet[see Figure 2 within]{Wood2018}.
 
We also computed the angular-momentum loss from $\lambda$~And, which amounts to $2.2 \times 10^{35}$ erg. While we have no good age estimates for $\lambda$~And, other than it is more evolved than the Sun, it has a much stronger spin-down rate than the Sun $\approx 2 \times 10^{5}\ \dot{J}_{\odot}$ (where $\dot{J}_{\odot} = 10^{29}-10^{30}$ erg,  \citealt{Garraffo2016,Finley2018}). It was previously believed that stellar rotation followed a simple power law with age \citep{Skumanich1972}, and this seems to hold true for main-sequence stars, but more recently it has been found that more evolved stars might not follow this relationship \citep{VanSaders2016,Booth2017, OFionnagain2019, Metcalfe2019}. More research into the complex relationship of stellar rotation, age and their activity is needed for evolved stars before we can accurately say what is happening here.

\begin{figure*}
    \centering
    \includegraphics[width=0.49\textwidth]{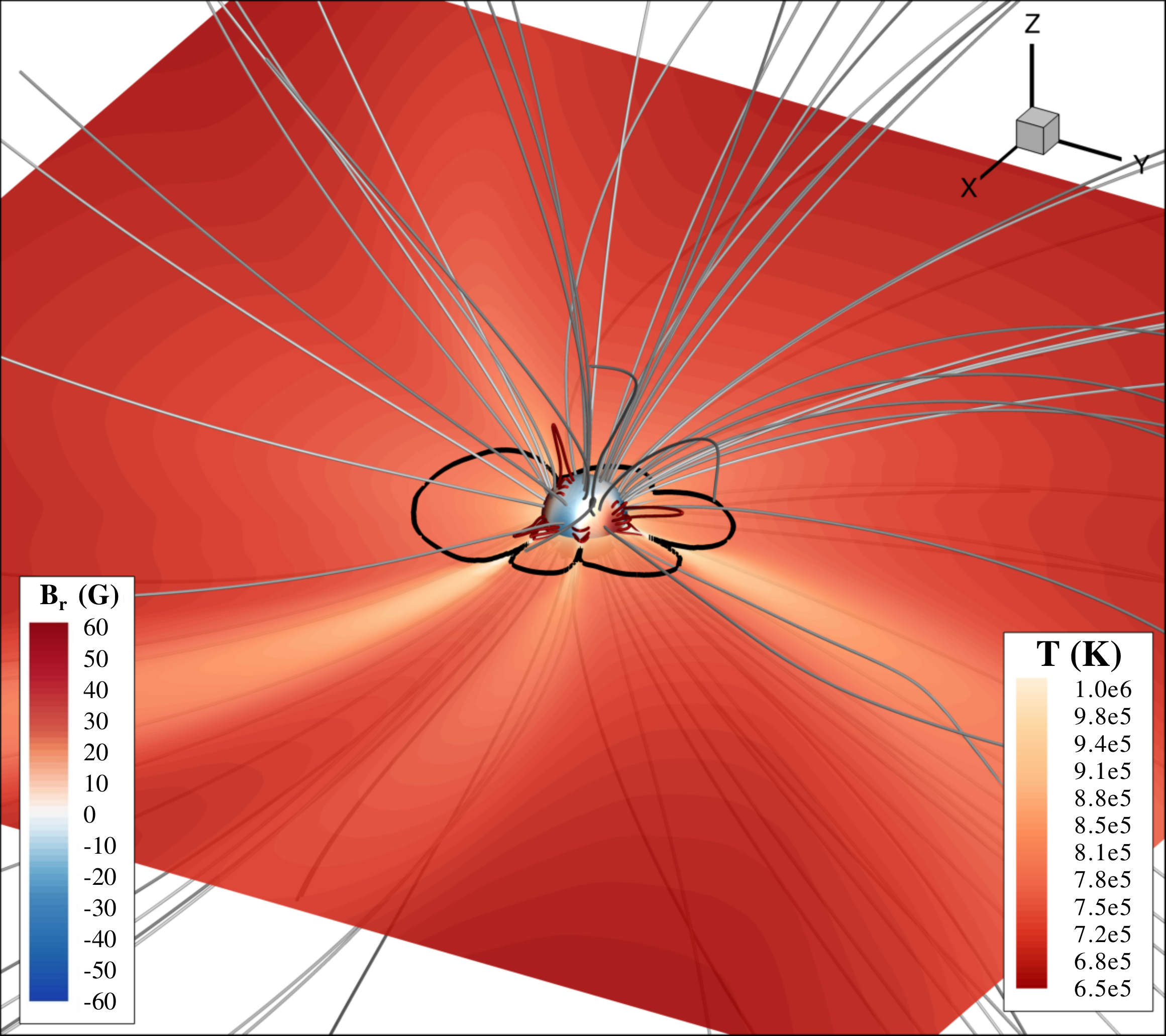}
    \includegraphics[width=0.49\textwidth]{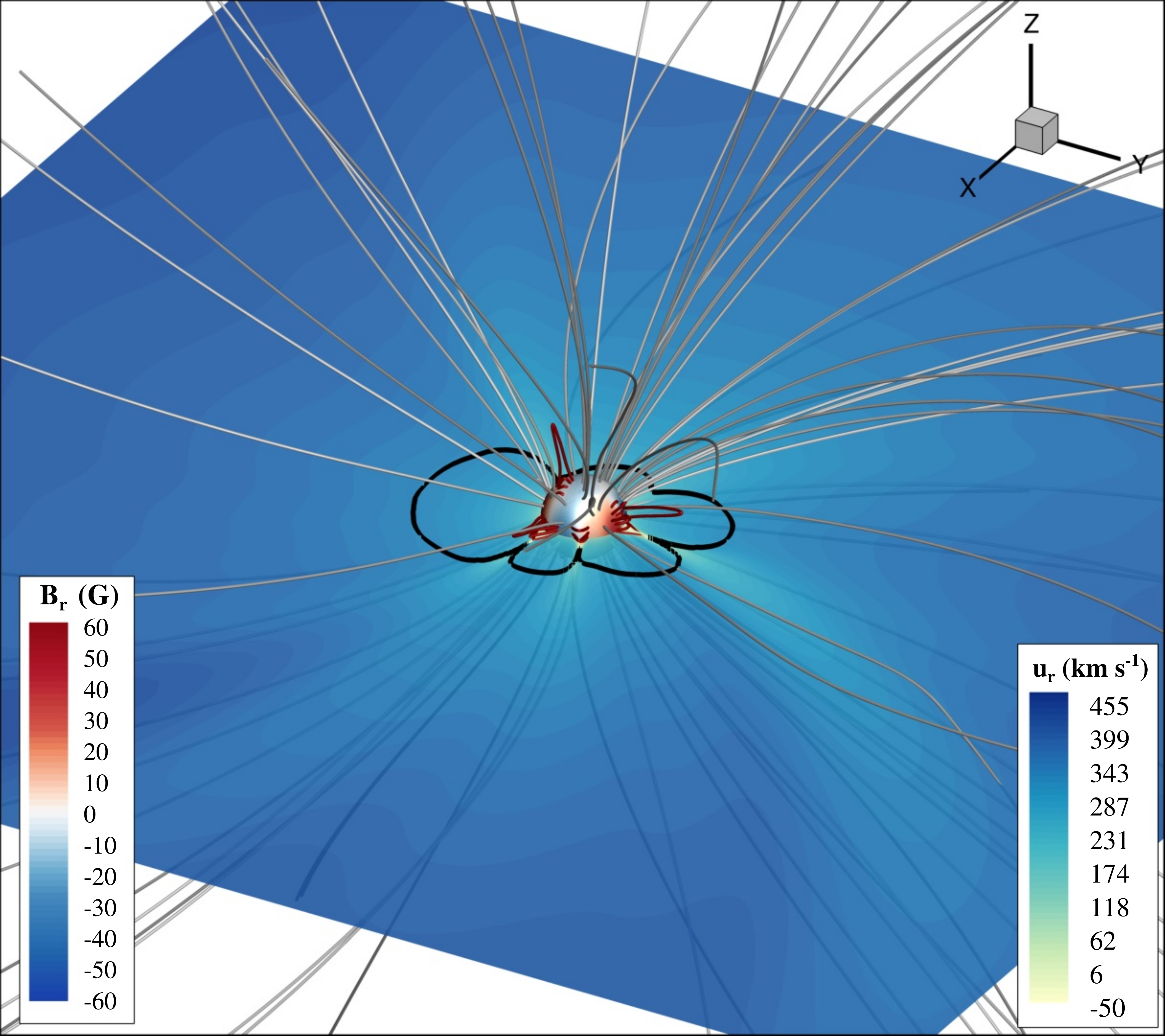}
  \caption{Result of our  3D MHD polytropic simulation. {Left:} Wind velocities displayed along the equatorial plane (yellow-blue) {Right:} Wind temperatures displayed along the equatorial plane (orange-red). In both plots, surface magnetic fields are displayed in blue-red. Closed magnetic field lines are red, open magnetic field lines are grey. The Alfven surface intersection with the x-y plane is shown as a black line. {The rectangular plane shown above extends from $-20$ to 20~$R_\star$ in each side.}}
  \label{fig:hotwind_tecplot}
\end{figure*}

{Our assumed base temperature for the polytropic wind model is 1MK, which is significantly smaller than the high temperatures of  $\sim$ 10 MK  seen in EUV and X-ray observations \citep{Sanz-Forcada2001,Sanz-Forcada2003}. This is because these high temperatures do not originate in the wind, but instead, similarly to the Sun, they are} believed to originate in small-scale magnetic field, likely due to reconnection/flaring events \citep{Priest2003, Aschwanden2004, Shibata2011, Lehmann2018}. In the Sun, the bulk of the high-energy emission (X-ray, EUV, for example) comes from regions of closed magnetic fields, while the solar wind comes from the coronal holes (open field lines) that are X-ray dark. This large-scale closed field line region is likely formed by a superposition of X-ray emission of small-scale flaring loops \citep{Vidotto2012}. Our ZDI map does not have the resolution to probe these small-scale loops, thus the X-ray/EUV properties derived in our models are underestimates of what would be the true X-ray/EUV emission coming from closed-field line regions. 

EUV observations suggest plasmas with temperatures $\approx$ 10 MK \citep{Sanz-Forcada2001,Sanz-Forcada2003}. In the framework of a polytropic wind model, a base wind temperature of this magnitude is much too large for such an extended stellar radius (which leads to low gravity). This results in the critical point existing inside the surface of the star and the wind begins supersonically, which is unphysical. From our 1D models, the largest possible base wind temperature is $\approx$ 1 MK. {The other free parameter of our model is the base density. Our  base density of $2.5\times 10^{10}~{\rm cm}^{-3}$ was chosen as it reproduces the observed radio flux densities of \obj . With this choice of base density, our radio flux density is $0.89$ mJy at a frequency of $4.5$ GHz, which is quite agreeable with an observed radio flux density of $\sim 0.8$ mJy at 4.5--5 GHz \citep{Bowers1981, Lang1985}, as we will discuss in details in the next Section. If, instead of this base density, we used} the derived plasma densities from EUV observations \citep[$2 \times 10^{12}$ cm$^{-3}$, ][]{Sanz-Forcada2001} as our base wind density, a 1D Parker wind predicts a mass-loss rate of $\approx 7 \times 10^{-7} \msano $ and a radio flux density of {$\approx 100$ mJy} at 5~GHz. This mass-loss rate is much larger than expected given the position of $\lambda$~And in the HR diagram (see \Cref{fig:cranmer_land}), and this radio flux density is $\approx 2$ orders of magnitude larger than observations (see \citealt{Bowers1981,Lang1985}). Therefore using the values from EUV measurements in our {polytropic wind} model would be unsuitable.

\subsection{Alfven-wave driven wind simulations}
Our second model, using the AWSoM code, results in relatively similar wind velocities, as can be seen in Table \ref{tab:AWSoM}, and a much cooler wind structure outside of magnetic loop regions. An example of an {Alfven-wave driven wind} model with a ZDI map at the lower boundary is shown in Figure \ref{fig:coldwind_tecplot}, and another one with a dipolar field as the lower boundary condition is shown in Figure \ref{fig:coldwind_tecplot_dipole}. For our simulations, we varied the {Poynting flux-to-magnetic ratio ($S_A/B$), the scaling for the correlation length ($\ell$)}, and the base wind density (n$_{\rm chr}$). $S_A/B$ alters the amount of energy the Alfven waves begin with at the base of the simulation, which can then be dissipated into the wind. $\ell$ changes the {correlation} length of the waves, increasing this value will cause the dissipation of energy to be much more extended, a small $\ell$ value will cause much of the energy to be deposited lower in the wind, near the chromosphere. 
{In the longitudinally-averaged temperature profiles shown in Figure \ref{fig_temperature}, for example, we see that the temperature starts with a small gradient, taking about 1 stellar radii above the surface to start increasing. The two wave cases presented in Figure \ref{fig_temperature}, in particular, have the largest correlation length-scale $\ell$ of our simulations.} In most of our Alfven-wave driven wind simulations, we find that the wind does not form a transition region similar to that of the solar case, in which the temperature increases from a few thousand K to a million K in a very short spatial scale \citep{2009A&A...501..745Y}. 

In terms of maximum temperature reached in our simulations, for} the case of $\lambda$~And, the density is high so radiative cooling dominates, hindering the formation of a hot wind. The base density plays a large role in the final wind structure as many physical processes depend heavily on the density structure. The wave dissipation (\Cref{eq:dissipation}), and particularly the radiative cooling (\Cref{eq:cooling}) have strong dependencies on density, and subsequently have strong consequences for the density structure in the wind. \citet{Suzuki2013a} (Figure 5 within) have shown how increasing input Poynting flux at the stellar surface can change the transition region height, and also cause a significant reduction in density with height in the wind. We see a similar effect in our simulations, which is consequential for our predicted radio flux densities. Our Alfven-wave driven wind  models all have substantially smaller radio fluxes than what is observed, as can be seen in \Cref{tab:AWSoM}. We will detail these results in \Cref{sec:radio}. We also note that while the {polytropic wind} solution discussed above does not present the high temperatures seen in X-ray observations of hot coronal lines, the Alfven-wave driven wind solutions  produce much hotter, albeit confined, regions within closed magnetic loops. Therefore the {Alfven-wave driven winds} produce hotter maximum temperatures as can be seen in \Cref{tab:AWSoM}. This is in better agreement with X-ray-derived temperatures from observations than that of the {polytropic wind} model. 

{Although  we have not directly computed the X-ray luminosities from our simulations---through radiative transfer methods like our work in \Cref{sec:radio} for radio emissions -- we can still compare the emission measures predicted in our simulations. For the Alfven-wave driven wind models \textbf{A1}, \textbf{D1} which represent the extremes of chromospheric density in our simulations, we estimated the emission measure which is defined as EM = $\int n_e n_i dV$, where $V$ is the volume where the integral is performed. The emission measure is an important quantity when considering higher energy emissions such as EUV and X-ray emission. We limit the calculation of EM from plasma above 1 MK by blanking regions of the simulation that possess temperatures colder than this. We find that the lower density model \textbf{A1} exhibits EM = $1.5 \times 10^{53}$ cm$^{-3}$, and the highest density model \textbf{D1} exhibiting EM = $6.3 \times 10^{48}$ cm$^{-3}$.\footnote{{Note that, although model \textbf{A1}  has a lower base density, it shows a higher EM because its high-temperature regions occupy a larger volume of the grid. Additionally, for the higher density model \textbf{D1},  the region where the temperature is above the assumed threshold in our EM calculations (1 MK) only starts at about $2 R_\star$ (Figure \ref{fig_temperature}), where the density has already dropped significantly. These two processes contribute to a higher EM in model \textbf{A1} than in model \textbf{D1}.}} We find EM that are similar but somewhat on the lower range of values found by \citet{Judge1986}. This is likely because our simulations miss the magnetic field at smaller scales, where most of the high energy emission is expected to originate. Therefore our estimated values are presented as lower-limits.}

\begin{figure*}
    \centering
    \includegraphics[width=0.49\textwidth]{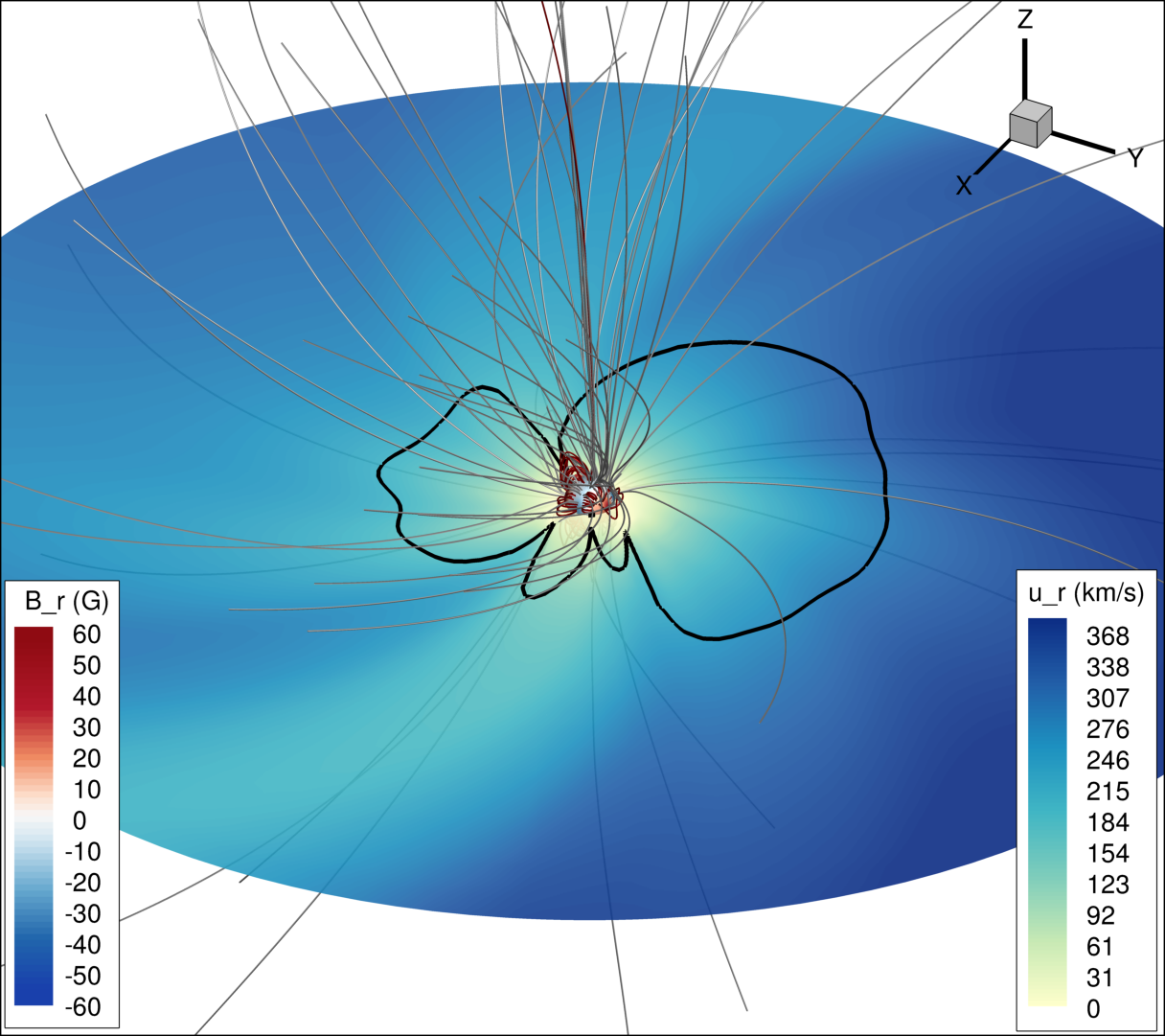}
    \includegraphics[width=0.49\textwidth]{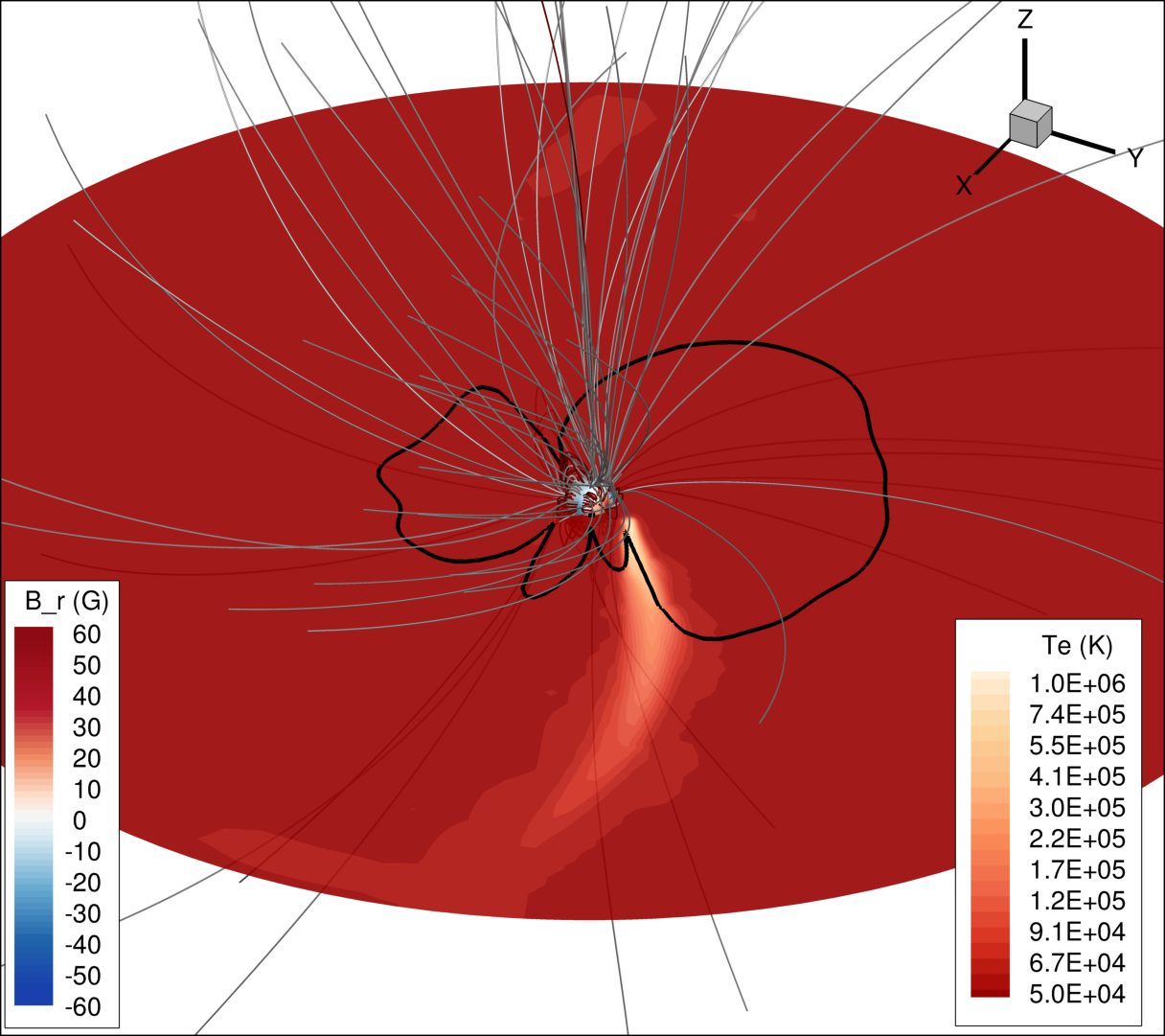}
  \caption{Model \textbf{C1}, one of the {Alfven wave-driven wind simulations}, with the ZDI map set as the inner boundary for the magnetic field, as in \Cref{fig:hotwind_tecplot}. {The circular plane shown above extends from $-30$ to $30~R_\star$ across its diameter.}}
  \label{fig:coldwind_tecplot}
\end{figure*}

\begin{figure*}
    \centering
    \includegraphics[width=0.49\textwidth]{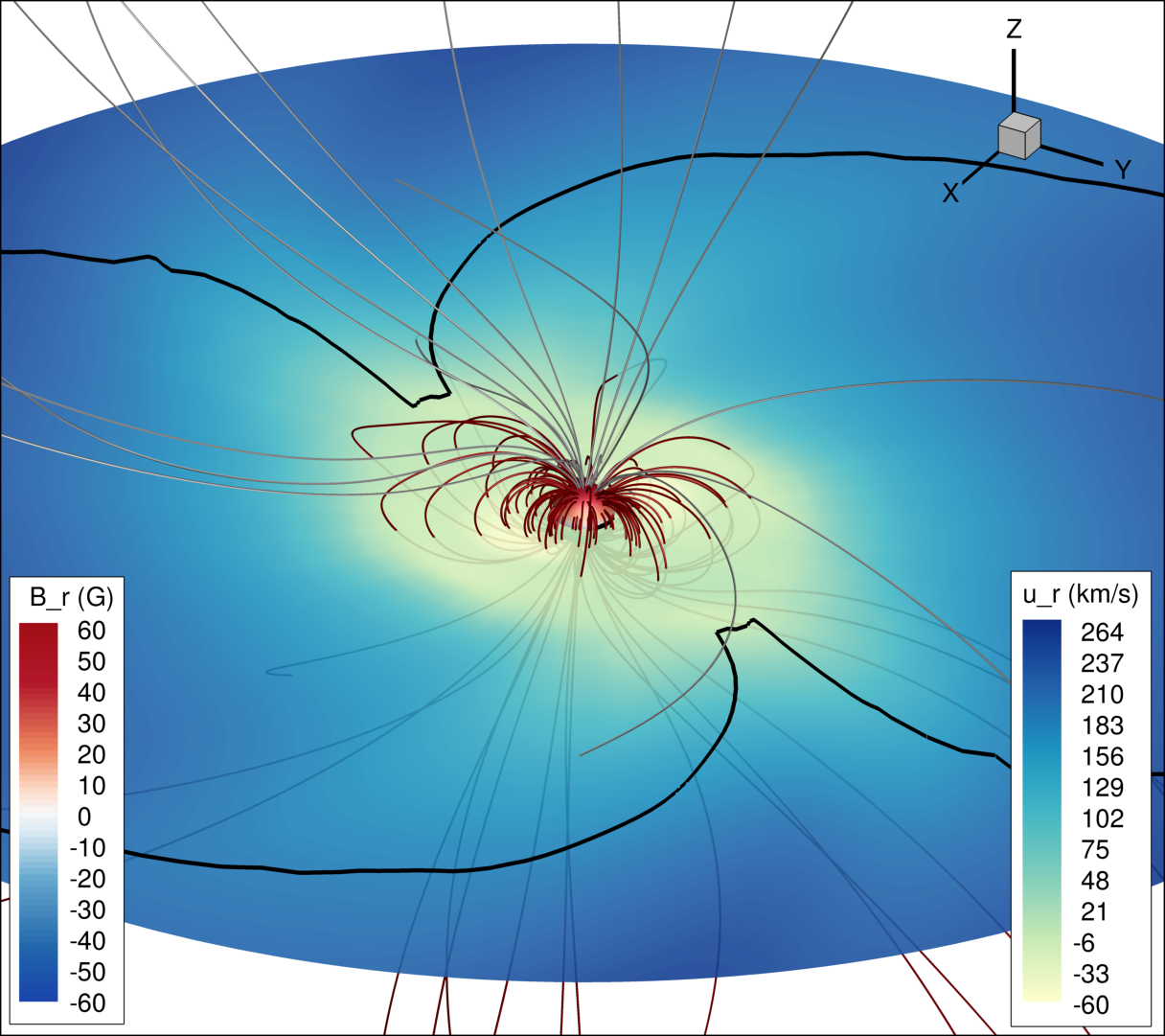}
    \includegraphics[width=0.49\textwidth]{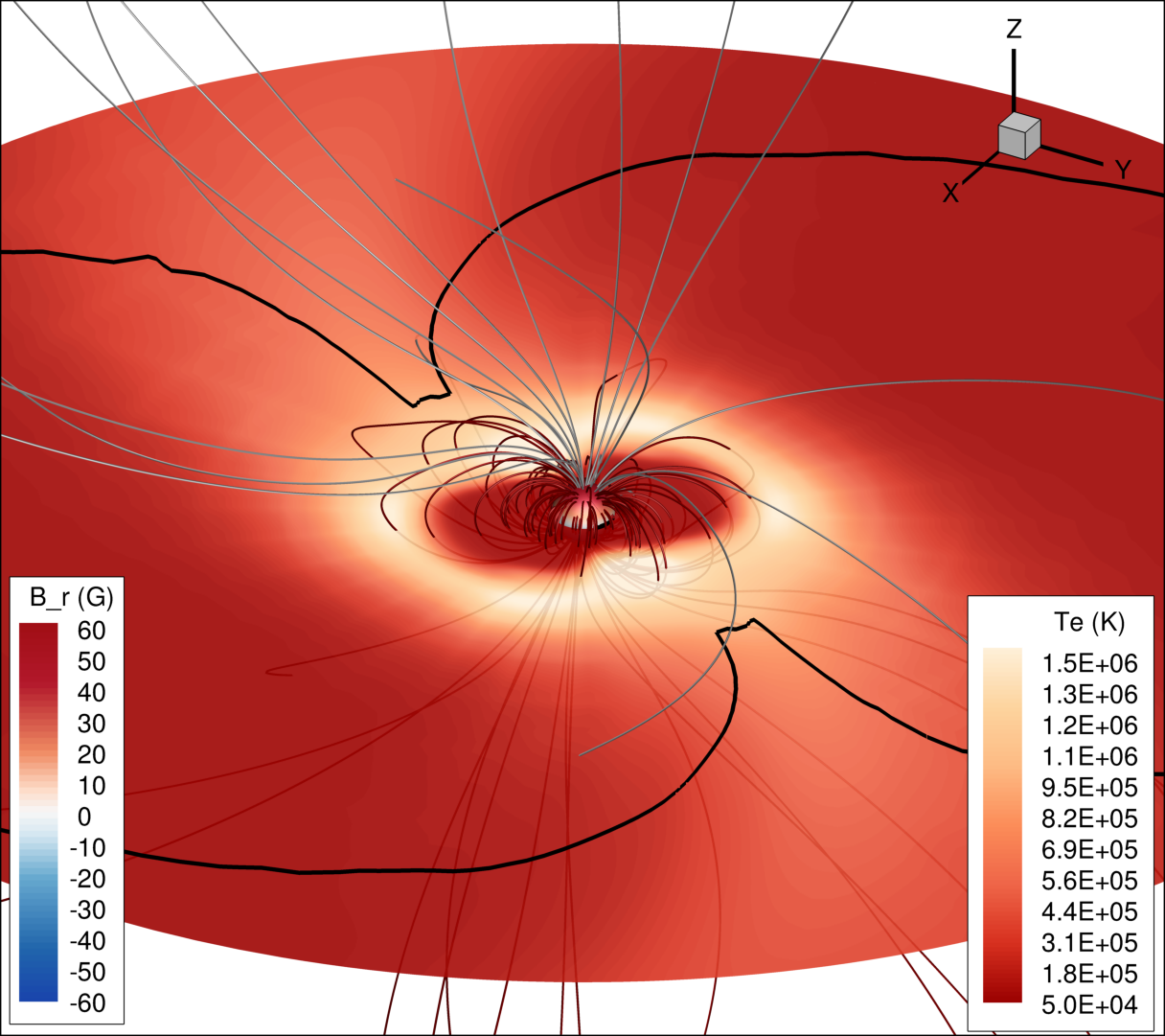}
  \caption{Model \textbf{D1}, one of the {Alfven wave-driven wind simulations}, with a dipolar field of B$_{\rm dip, max} = 60 G$, as in \Cref{fig:hotwind_tecplot}.  {The circular plane shown above extends from $-30$ to $30~R_\star$ across its diameter.}}
  \label{fig:coldwind_tecplot_dipole}
\end{figure*}

\begin{figure}
    \centering
    \includegraphics[width=\columnwidth]{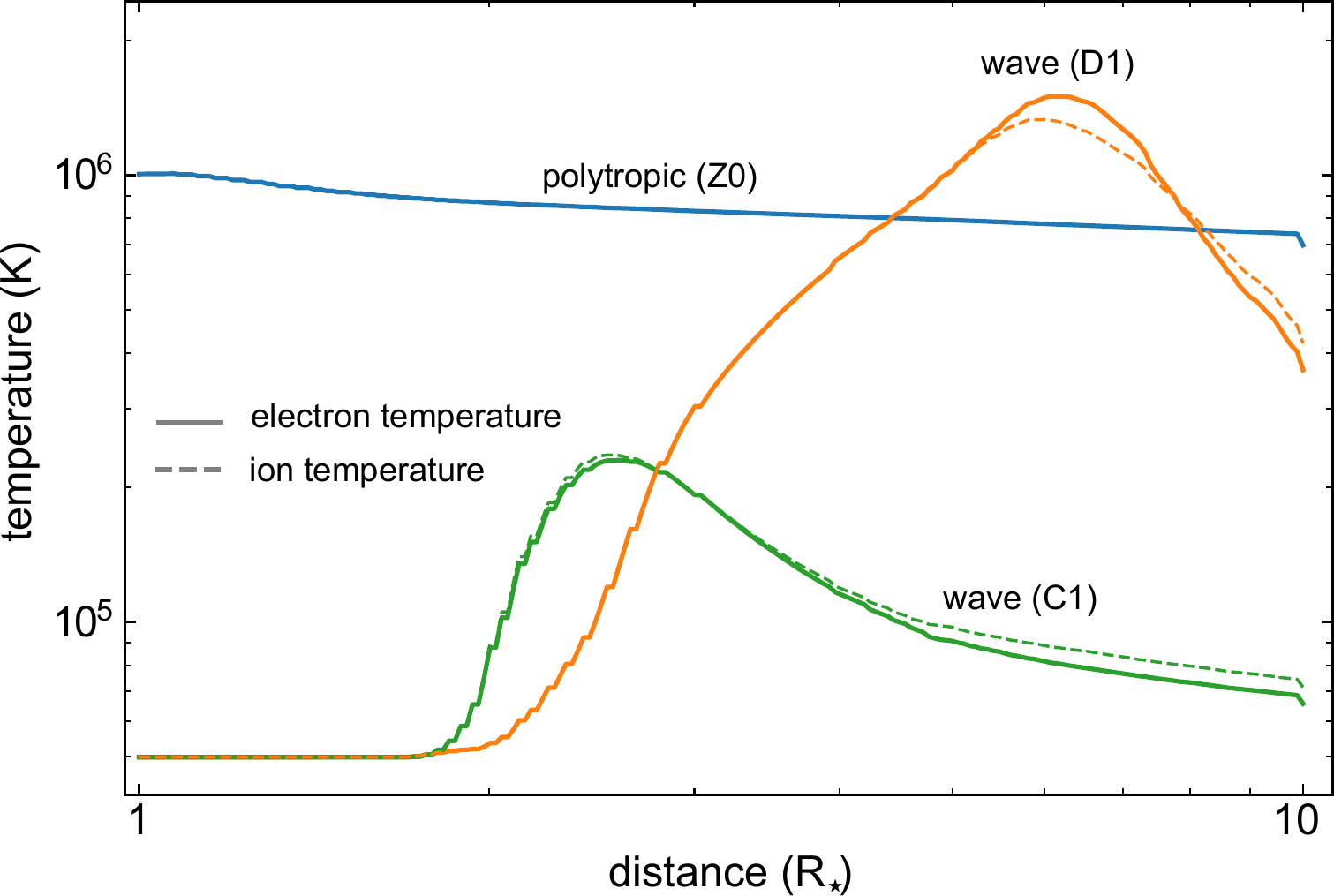}
    \caption{{The averaged temperature around the equator of 3 simulations shown in blue (polytropic, \Cref{fig:hotwind_tecplot}), green  (wave-driven \textbf{C1}, \Cref{fig:coldwind_tecplot}), and orange  (wave-driven \textbf{D1}, \Cref{fig:coldwind_tecplot_dipole}), respectively. Solid lines are the electron temperature, while the ion temperature is shown by the dashed lines.}}
    \label{fig_temperature}
\end{figure}

Our Alfven-wave driven wind models produce much lower mass-loss rates than our {polytropic wind} models. The largest mass-loss rate from our Alfven-wave driven wind models is $1.9 \times 10^{-10} \msano $, which is still one order of magnitude less than our {polytropic wind} model. The lowest mass-loss rates calculated come from the \textbf{A4} model, with $5.6 \times 10^{-13} \msano $. This is one order of magnitude larger than the currently accepted solar mass-loss rate. We find it difficult to constrain this value for these simulations as these winds do not produce similar radio flux densities to the observed values (see \Cref{sec:radio}). The Alfven-wave driven wind model produces slightly lower angular momentum-loss rates than the {polytropic wind} model. Note that in the case of these simulations, the uncertainties in the magnetic field, or differences in magnetic field geometry do not dominate the uncertainties in the global wind parameters. The uncertainties in the Poynting flux and damping length parameters are more important.

Figures \ref{fig:hotwind_tecplot}, \ref{fig:coldwind_tecplot} and \ref{fig:coldwind_tecplot_dipole} show the Alfven surfaces (the surface where the wind velocity equals the Alfven velocity) as black contours. While the Alfven surfaces in the {polytropic wind} (\textbf{Z0})  and Alfven-wave driven wind (\textbf{C1}) models are relatively small, we see that the Alfven surface is quite extended in the dipolar model \textbf{D1} ($\approx 30$ $R_{\star}$). However, this large size is only an inclination effect. Dipolar fields produce Alfven surface with a dumbbell shape (e.g. the case of GJ49 in \citealt{Vidotto2014b}). In model \textbf{D1}, this dumbbell surface extends from $\approx$ 5 $R_{\star}$, out to $\approx$ 30 $R_{\star}$. Because of the dipolar tilt of the magnetic field, the dumbbell shape is tilted, crossing the equatorial plane at 30 $R_{\star}$, this extends beyond the orbit of the secondary star. {With a circular orbital period of 20.5212~d \citep{Walker1944}, the estimated orbital separation is $\simeq 23 R_\odot \simeq 4 R_\star$ \citep{Donati1995}.} In this case, interesting effects can take place in the system. Perturbations caused by an orbiting companion could travel downwind through plasma waves, allowing this information to reach the base of the wind and modifying the wind structure globally. This is similar to the physical processes seen in the cases of exoplanets orbiting in sub-Alfvenic regions \citep{Strugarek2019,Folsom2020}. We ignore the companion star in this work and assume that the companion is not actively affecting the stellar wind, which might not be true, in the simulation case \textbf{D1}, for example. Of course in our other simulations the Alfven surface is much less extended, in which case, we expect the companion not to affect the stellar wind. 

\section{Using radio emission of \texorpdfstring{\boldmath$\lambda$}{lambda} And to select the most appropriate wind model}\label{sec:radio}
We have a number of observational constraints that we can use as a guidance for selecting the most appropriate model that describe the wind of \obj , such as the X-ray derived temperatures and emission measures (see last Section). The mass-loss rate is another parameter that can be used to select the most appropriate model for the wind of \obj . For example, the location of $\lambda$~And on the HR diagram provides loose constraints on mass-loss rates based on studies of other evolved low-mass stars: $10^{-11}$--$10^{-9} \msano $  (see \Cref{fig:cranmer_land}, \citealt{Cranmer2019}). It is possible that $\lambda$~And is transitioning from a hot corona to no corona implying it could have a ``hybrid'' wind, with mixed characteristics of the hot coronal winds and the cool wave-driven winds. However, the X-ray observations point more strongly towards $\lambda$~And still showing signs of a hot corona  \citep{Ortolani1997,Sanz-forcada2004,Drake2011}, with maximum coronal temperatures of $7-10$ MK. While giants more evolved than $\lambda$~And usually possess winds with low terminal velocities (< 40 km/s, \citealt{Drake1986, OGorman2018}) or even slightly faster ($\lesssim 150$ km/s, \citealt{2009AJ....138.1485D}), the presence of a hot corona is likely to lead to moderate terminal wind velocities ($\approx300-400$ km s$^{-1}$), and mass-loss rates of $10^{-11}$--$10^{-9} \msano $ \citep{Linsky1979,Drake1986}. The aforementioned mass-loss rate derived from comparison to neighbour stars in the HR diagram, however, is at odds with the mass-loss rate derived in astrospheric observations, which can be as low as $2 \times 10^{-15} \msano $ and as high as  $1 \times 10^{-13} \msano $ \citep{Muller2001,Wood2018}.

The several orders of magnitude differences in the mass-loss rates of $\lambda$~And derived so far in the literature has led to us to use a different approach to constrain the wind of $\lambda$~And, {namely} using radio observations. In Section \ref{sec:intro} we discussed how stellar winds can be constrained or detected through radio observations  \citep{Panagia1975, Wright1975,2002ARA&A..40..217G}. This thermal radio emission scales with the wind plasma density squared ($\propto n^2$), which means the tenuous winds of low-mass main sequence stars remain mostly undetectable for current radio telescopes. However, in the case of solar-mass red giants, their winds are much denser due to an increase in mass-loss rate \citep{2016ApJ...829...74W}, allowing these winds to be readily detected at radio wavelengths. As a result, radio emission from stellar winds provide us a direct detection of the wind, which limits the {base} density adopted in our simulations and, consequently, the mass-loss rate \citep{Vidotto2017a, OFionnagain2018, OFionnagain2019}.

\subsection{Radio observations}\label{sec.obs_radio}
One of the goals of our work is to use radio observations of $\lambda$~And to constrain our wind models. A compilation of the observed radio emission for \obj\ is shown in Table \ref{tab_radio}. As can be seen, radio observations of $\lambda$~And were mainly published a few decades ago \citep{Bath1976,Bowers1981,Lang1985}.  We queried the  National Radio Astronomy Observatory (NRAO) data archive for prior unpublished observations of \obj\ and recovered a number of more recent observations with the VLA. In some cases, multi-band data were available from a single epoch, allowing us to reproduce broadband spectral energy distributions. These data are key to identifying one or more components of radio emission from \obj\ and are the focus of our analysis here. 

We detect a large flux density on 1998 Dec 12 (see Figure \ref{fig_VLA}) that is almost certainly non-thermal emission and not caused by the stellar wind. In addition to this, we see a strong evidence for a basal quiescent flux level at around $0.5$ -- $0.6$~mJy at 4.75~GHz, detected in 1998 Dec 17 and 1999 Feb 13. Our quiescent emissions are in line with  values reported in \citet{Bowers1981} and \citet{Lang1985}. These quiescent components could have a thermal origin (e.g., from the stellar wind), but could also be gyrosynchroton. We discuss how this affects our conclusions in Section \ref{sec.thermal}.

\begin{table}
    \centering
    \caption{{Compilation of radio observations of $\lambda$~And. We show each flux density for observing frequencies. Literature values are shown at the top part of the table, and our newly reported observations of \obj\ with VLA are shown at the bottom part of the table. VLA observations are from program AN0083 (1998 Dec 12 and 17) and from program AW0362 (1999 Feb 13).}}
    \begin{tabular}{ccl}
    \hline
    $\nu$ [GHz] & $\Phi_{\rm radio}$ [mJy] & Reference \\ \hline \hline
    2.72 & $< 15$ & \citet{Bath1976}\\
    5 & 65 & \citet{Bath1976}\\
    8.1 & 20 & \citet{Bath1976}\\
    5 & 0.84 & \citet{Lang1985} \\
    4.5 & 0.86 & \citet{Bowers1981}\\ \hline
    4.75 &    $37.6 \pm 3.8$ & 1998 Dec 12, this work\\
   8.26 &  $27.7 \pm 2.8 $ & 1998 Dec 12\\
    14.9 &   $16.7 \pm 1.7$ &   1998 Dec 12\\
    
    4.75 &    $0.53 \pm 0.08$ & 1998 Dec 17\\
   8.26 &  $0.37 \pm 0.06 $ & 1998 Dec 17\\
   14.9 &   non detection &  1998 Dec 17\\
    
     4.75 &    $0.63 \pm 0.09$ &  1999 Feb 13\\
   8.26 &  $0.73 \pm 0.09 $ & 1999 Feb 13\\
   14.9 &   non detection &   1999 Feb 13\\
       
    \hline
    \end{tabular}
    \label{tab_radio}
\end{table}

\begin{figure*}
    \centering
    \includegraphics[width=0.8\linewidth]{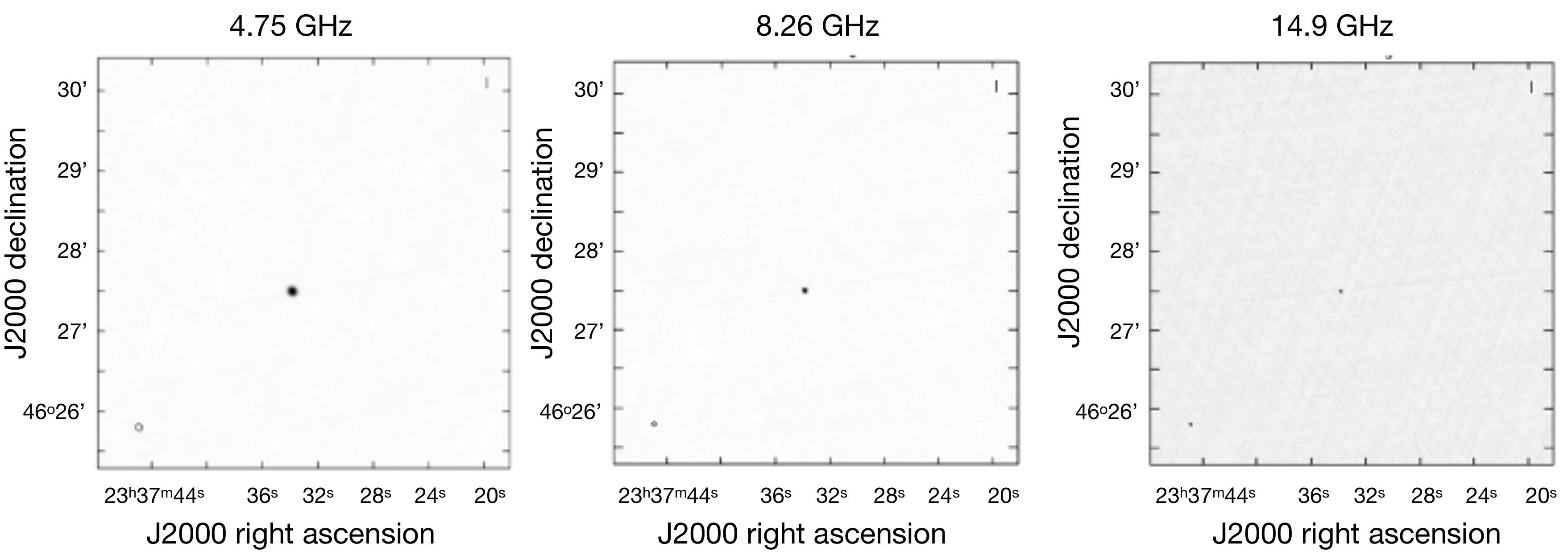}
    \caption{{VLA observations of \obj\ from 1998 Dec 12 (program AN0083) at the C, X, and U bands (4.75, 8.26, 14.9 GHz), during an active state of the star.}}
    \label{fig_VLA}
\end{figure*}

\subsection{Comparing simulations and observations}
{The polytropic wind scenario shows a radio flux density of $0.89$ mJy at a frequency of $4.5$ GHz, which is very similar to the observed values of $\sim 0.8$~mJy at 4.5--5 GHz \citep{Bowers1981, Lang1985} and our newly derived quiescent values ($\sim 0.5$ -- $0.6$~mJy at 4.75~GHz) from 1998 Dec 17 and 1999 Feb 13. This is due to our choice of base parameters for this model, such as the base density. The {specific radio intensity $I_\nu$ of the polytropic} wind model is shown in the left panel of \Cref{fig:radiointensity}. We can see that there is quite an extended region of specific radio intensity, outside of the optically thick region, delineated by the dashed contour. The Alfven-wave driven wind models, on the other hand, predict lower-than-observed radio fluxes ($\lesssim 0.03$~mJy for all models shown in Table \ref{tab:AWSoM}). The middle and right panels of Figure \ref{fig:radiointensity} show the specific radio intensity for two Alfven-wave driven wind models (\textbf{C1} and \textbf{D1}, respectively), where we see much lower radio intensities in the optically thin region compared to the polytropic model.}

The geometry of the optically thick region, including its size, is determined almost exclusively by the density structure of the wind. We find that in the case of the Alfven-wave driven wind simulations, the density structure results in radio flux densities {$S_\nu$} that are very low and do not agree with observations.  In the case of the Alfven-wave driven wind, none of our simulations reached the required level of radio flux to reproduce the {quiescent levels seen in the observations (compare Tables \ref{tab:AWSoM} and \ref{tab_radio})}. The largest radio flux of any of the Alfven-wave driven wind simulations was the \textbf{C2} case, which possesses a very dense chromospheric boundary condition of $1.5 \times 10^{13}$ cm$^{-3}$. The wave-driven wind, due to the cold inner regions that are almost isothermal {(see Figure \ref{fig_temperature})}, causes a strong exponential density decay with height. 
We found that increasing the base density further does not necessarily increase the radio flux, as the density drops off in a more rapid fashion than the lower base density cases. This happens as the high density near the star causes this region to become hydrostatic-like, leading to an exponential decay in wind density. The higher base density, being so confined near the star, does not contribute much to the emitted radio flux density at $4.5$ GHz as the wind is optically thick out to $\approx2$ $R_{\star}$. As a result of the exponential decay, outside this optically thick region, the wave-driven winds display much lower density than the thermally driven polytropic wind, resulting in lower radio flux densities. 
 
\begin{figure*}
    \includegraphics[width=0.33\textwidth]{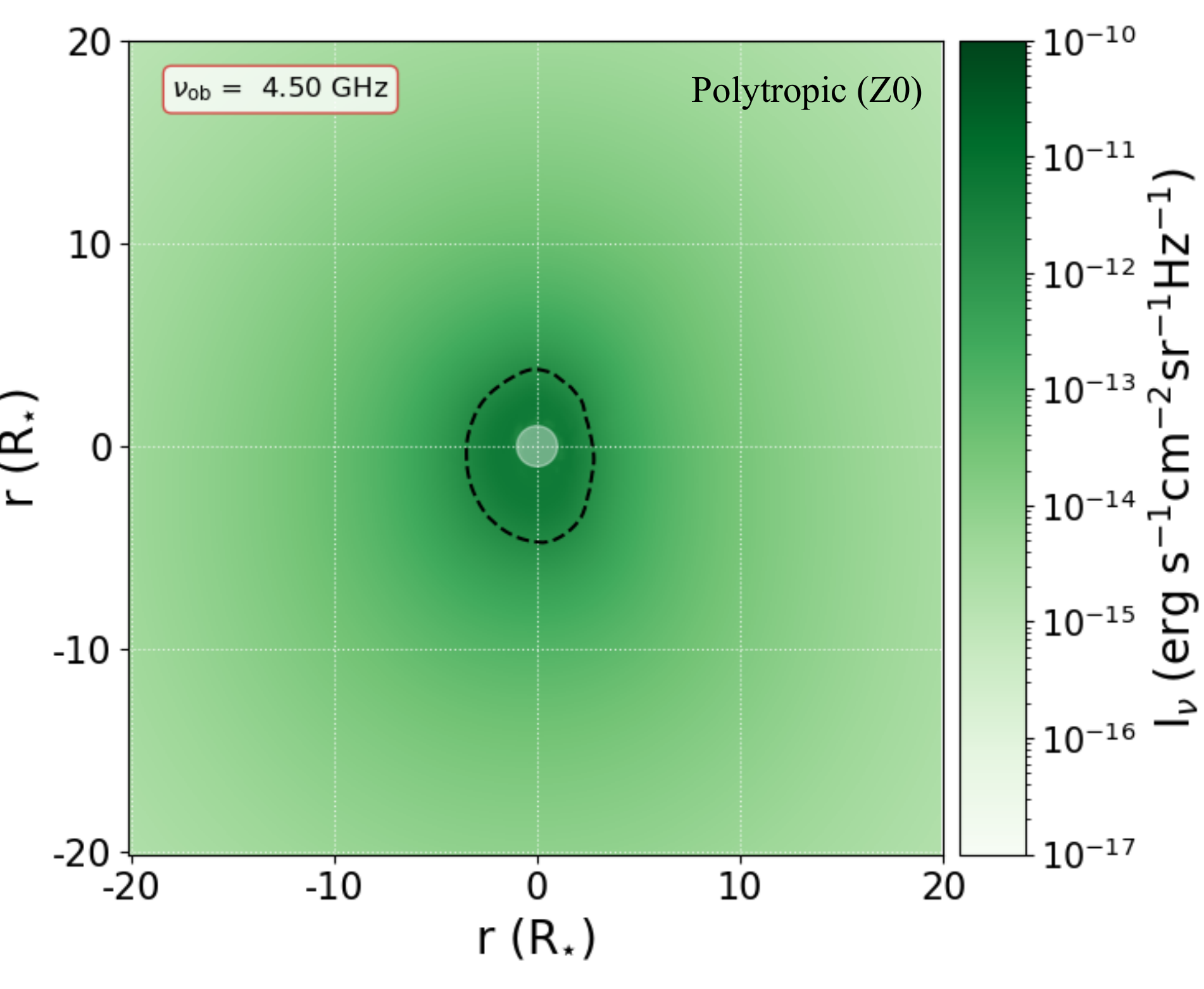}
    \includegraphics[width=0.33\textwidth]{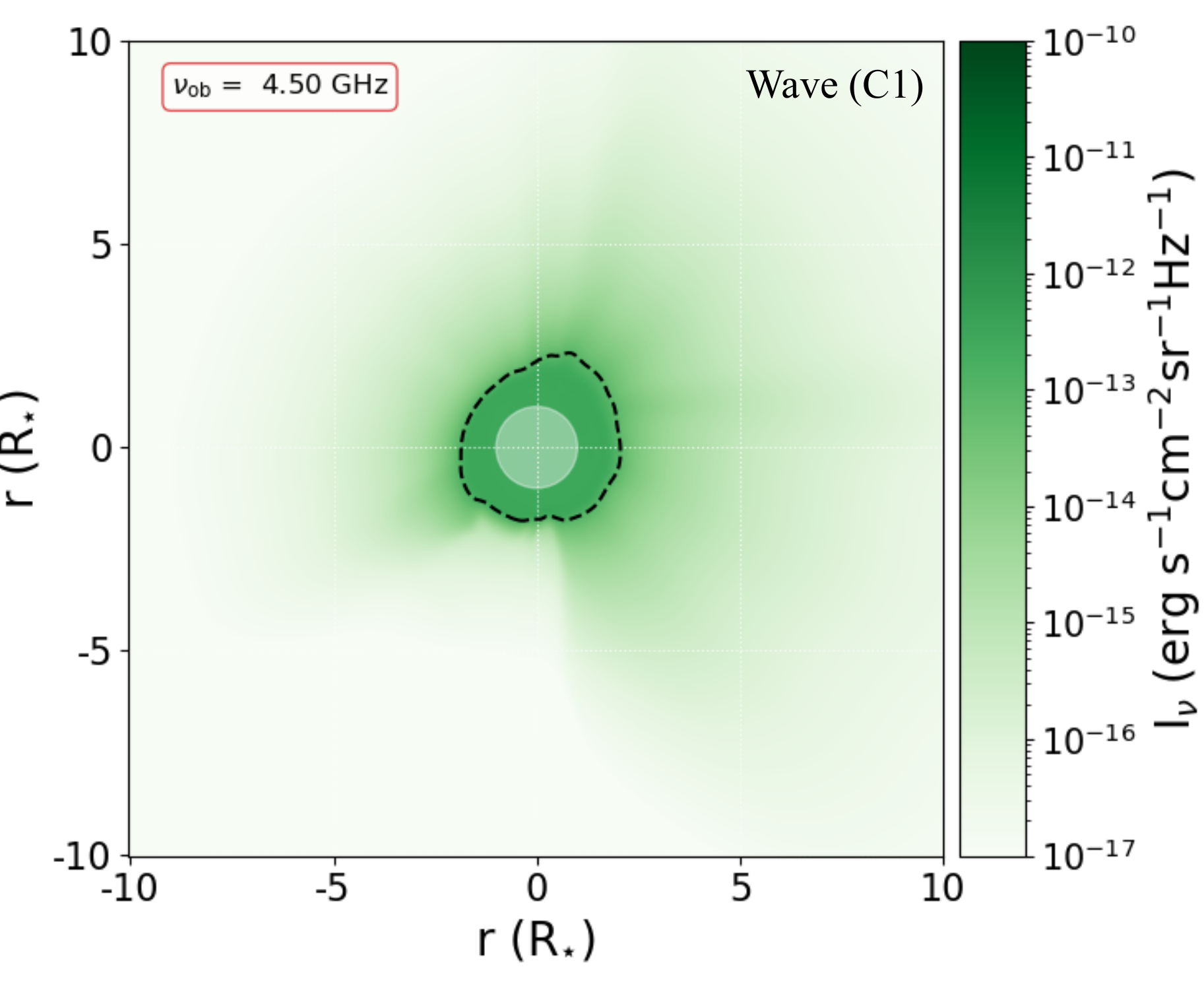}
    \includegraphics[width=0.33\textwidth]{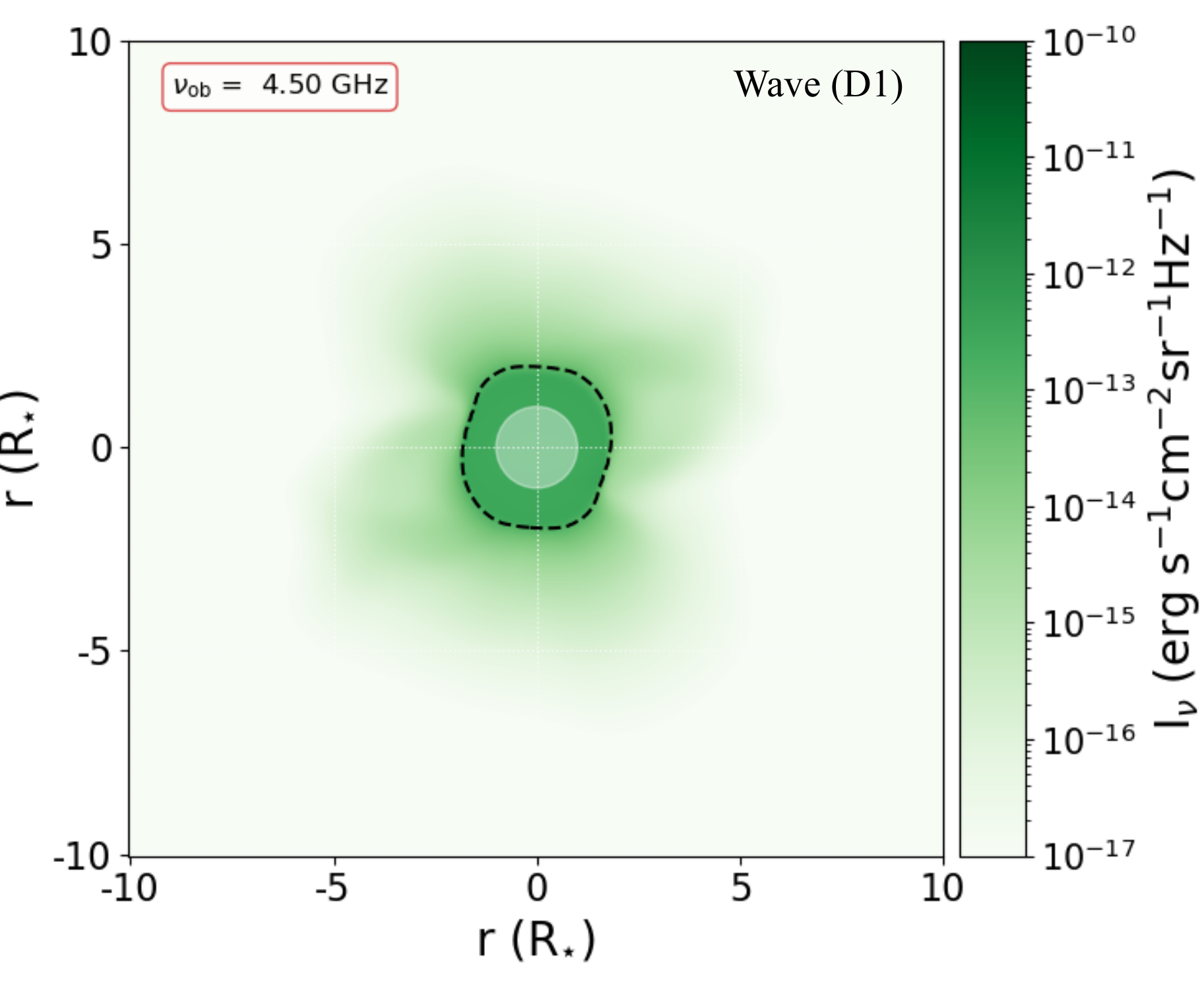}
  \caption{Radio intensities of the {polytropic wind}, Alfven-wave driven wind with a ZDI map, and Alfven-wave driven wind with a dipole, respectively, shown in \Cref{fig:hotwind_tecplot,fig:coldwind_tecplot,fig:coldwind_tecplot_dipole}. Plots from the Alfven-wave driven wind models are zoomed in to $[-10, 10]~R_\star$ to display more detail. It is evident from these plots that the density decay in the Alfven-wave driven winds has a significant effect on the radio flux density emitted from the wind. {The black dashed contour shows the optically thick surface of $\tau = 0.399$ at 4.5~GHz, which delineates the region within which half of the emission originates \citep{Panagia1975}.}}
  \label{fig:radiointensity}
\end{figure*}

This is more easily illustrated in \Cref{fig:density}, which shows the averaged equatorial density profile for the three plotted simulations in \Cref{fig:hotwind_tecplot,fig:coldwind_tecplot,fig:coldwind_tecplot_dipole}. The {polytropic wind} model (shown in blue) begins with a lower base density, but with a mostly $r^{-2}$ dependency on distance. The Alfven-wave driven wind models however, produce an exponential decay in density up to $\approx 2$ $R_{\star}$, at which point, they have a much lower density than the {polytropic wind} scenario. {This exponential decay is similar to that of a hydrostatic, isothermal atmosphere, i.e., $\rho \propto \exp\{-(r-R_\star)/H_0\}$, where  $H_0 = k_B T / (m g)$ is the scale-height of a plane-parallel atmosphere, in which the temperature and gravity remains constant with altitude. 
To guide our eyes, the solid grey line in Figure \ref{fig:density} shows the exponential decay for a scale height of $H_0 \simeq 0.05R_\star$. This strong decay in density results in much lower radio intensity in the optically thin region for the Alfven-wave driven winds, as shown in the middle and right panels of Figure \ref{fig:radiointensity}.}

\begin{figure}
    \centering
    \includegraphics[width=\columnwidth]{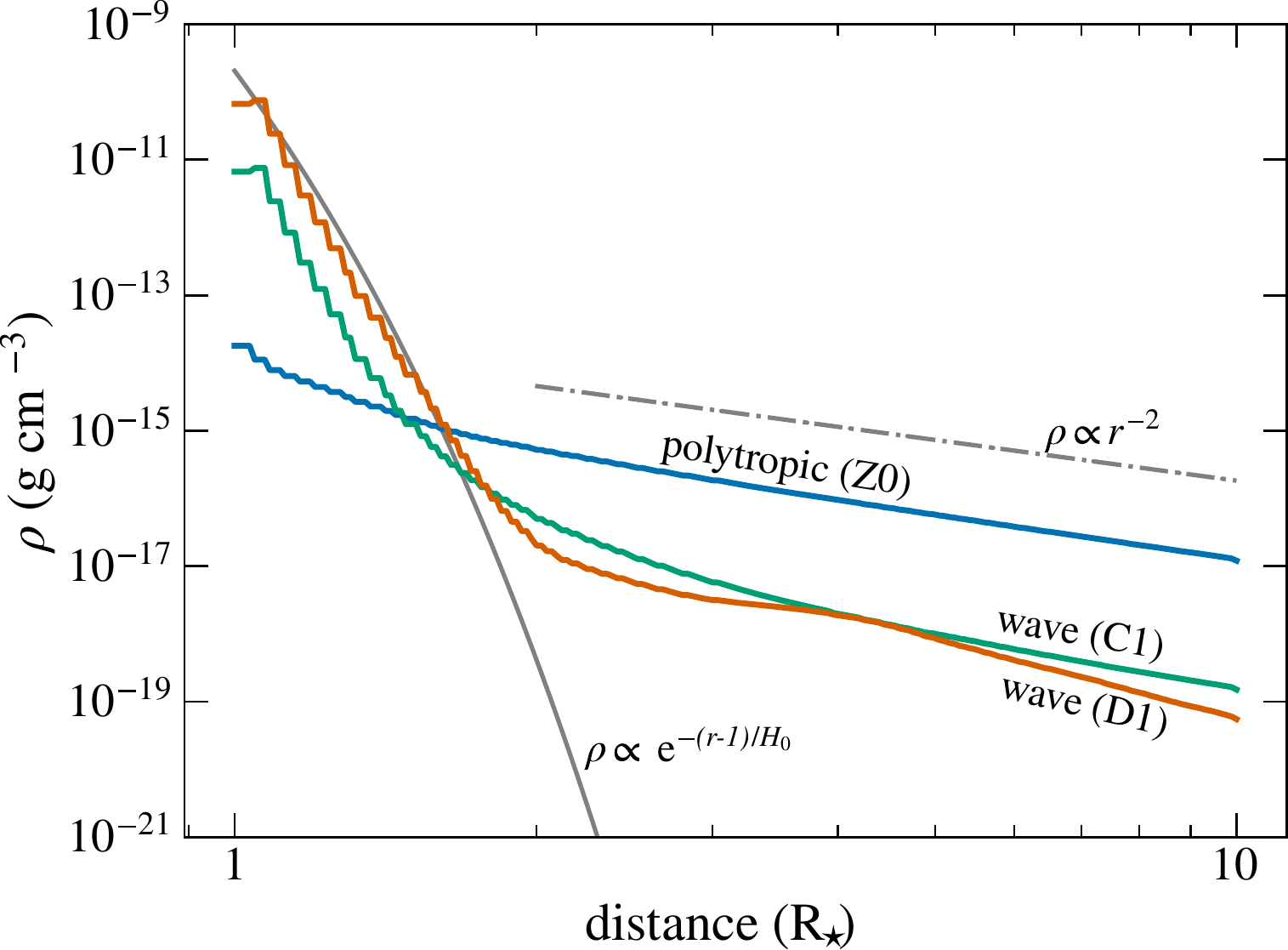}
    \caption{In this plot we show the averaged density around the equator of 3 simulations shown in blue (polytropic, \Cref{fig:hotwind_tecplot}), green (wave-driven \textbf{C1}, \Cref{fig:coldwind_tecplot}), and orange (wave-driven \textbf{D1}, \Cref{fig:coldwind_tecplot_dipole}) respectively. The  {Alfven-wave driven winds} display hydrostatic behaviour close to the stellar surface and we see an exponential decay with a scale height of $H_0 =  0.05~R_{\star}$. Gray lines show examples of r$^{-2}$ (dot-dash) and exponential decay (solid).}
    \label{fig:density}
\end{figure}

\subsection{Thermal versus non-thermal radio emission}\label{sec.thermal}
By comparing our models with radio observations of \obj , we have an opportunity to constrain  stellar wind parameters, such as its mass-loss rate. However, the interpretation depends on the nature of the quiescent radio emission from \obj , namely whether it is thermal (from the stellar wind) or non-thermal (gyrosynchrotron). 

If the observed radio emission of \obj\ is thermal and originates from its wind, we are able to place a {\it firm} constraint on the density generating this emission and thus on the mass-loss rate of the wind of \obj . In this case, the better agreement between our models and the quiescent values of radio flux densities ($\sim 0.8$~mJy at 5~GHz) would lead us to conclude that the polytropic, thermally-driven wind is likely to be the best description of the wind of \obj\ and that this results in a well-constrained mass-loss rate of $3 \times 10^{-9}\msano$. This is in line with mass-loss rates of neighbouring stars in the HR diagram (\Cref{fig:cranmer_land}), but much higher than values predicted by the astrosphere method \citep{Wood2018}. 

However, the quiescent emission shown in Section \ref{sec.obs_radio} could be gyrosynchrotron emission (not wind). When this happens, the lower quiescent values cannot be used to constrain stellar wind models and we use, instead, the flaring, non-thermal emission of \obj\  (see, e.g., Figure \ref{fig_VLA} and Table \ref{tab_radio}). If we are able to detect this non-thermal emission, it means that the wind of \obj\ is likely optically thin, which allows radio flares to escape the system \citep{Lim1996, Fichtinger2017}.  In the scenario of non-thermal radio emission, our models can help us estimate the {\it maximum} density, and thus the {\it maximum mass-loss rate}, of \obj\ -- as any greater density would have absorbed the flaring emission of this system. In this case, our wind models indicate that the mass-loss rate of \obj\ is $\lesssim 3\times 10^{-9}\msano$, if this star has a thermally-driven wind. If the wind acceleration is caused by the dissipation of Alfven waves, then our upper limit in the mass-loss rate of \obj\ is one order of magnitude lower, i.e., $\lesssim 2 \times 10^{-10}\msano$. 

\section{Summary and Conclusions} \label{sec:conc}
In this work, we modelled the wind of the post-main sequence star \obj , which is a solar-mass, sub-giant star. For that, we performed spectropolarimetric observations, which allowed us to derive the large-scale surface magnetic field of \obj\ used as input in our models. Additionally, our wind models were constrained by the radio emission of this star that have been presented in previous works as well as newly reported  archival VLA data that we presented here.
 
BritePol spectropolarimetric observations from August to October 2016 of $\lambda$~And were obtained and used to derive a surface magnetic field through ZDI. We found a magnetic field that exhibits mostly low order spherical harmonics (78\% are $\ell \leq 3$), with most of the magnetic energy in the poloidal component. The geometry of the field is tilted at 90$^{\circ}$ with respect to the rotation axis. We found a maximum local magnetic field of $83$ G, with an unsigned average of $21$ G (\Cref{fig:zdimap}). This is a relatively strong magnetic field compared to the solar magnetic field, considering the evolved state of $\lambda$~And. 

Using the ZDI magnetic map, we carried out simulations using two different wind models: a {polytropic wind} model and a Alfven-wave driven wind model. We included here a single hot (1MK) polytropic wind case. In the wave-driven model, we run a set of simulations varying the input parameters of {Poynting flux-to-magnetic ratio ($S_A/B$) and the scaling for the correlation length ($\ell$), which is related to the turbulent damping of the waves}. We find that increasing Poynting flux consistently results in hotter, faster stellar winds, while damping length has a more complicated relationship to wind velocity and temperature, with shorter damping lengths depositing more energy near the star. 

The maximum temperatures we find exist in our Alfven-wave driven wind simulations (T$_{\rm max} = 11$ MK; model \textbf{A1}), but are notably confined to small regions in the wind. We see a maximum temperature of 1 MK in our {polytropic wind} model, which is markedly below the derived temperatures from X-ray observations. This is due to the lack of small-scale magnetic field in these simulations. It is generally accepted that the small-scale field, which can produce strong local magnetic fields, produces the hottest plasma, which emits hot X-ray lines. The ZDI technique is not sensitive to these small-scale fields, and so they are excluded from our simulation. Furthermore, the stellar winds emanate from open field regions, whereas it is the closed field regions that produce this hot plasma. 

In our simulations, we are able to calculate mass-loss rates $\dot{M}$ and angular momentum-loss rates $\dot{J}$, the latter of which show similar values for all our models. Our {polytropic wind} simulation displays a strong spin-down rate of $\dot{J} = 2.2 \times 10^{35}$ erg. Our Alfven-wave driven wind model maximum spin-down rate is similar at $\dot{J} = 1.1 \times 10^{35}$ erg for a dipolar magnetic field, and $\dot{J} = 4.5 \times 10^{34}$ erg for the observed surface magnetic field. As there is no consensus on low-mass stellar spin-down rates for stars older than our Sun, it is difficult to place these spin-down rates in a solar evolutionary context, but these angular momentum-loss rates are much larger than current accepted values for the solar wind.

Although all our models show similar angular momentum loss rates, our derived mass-loss rates are quite different, depending on the physics of the wind models. The mass-loss rate of our {polytropic wind} model is $ 2.9 \times 10^{-9} \msano $, while the Alfven-wave driven wind model produces lower mass-loss rates overall, with a high of $1.7 \times 10^{-10} \msano $ and a low of $5.6 \times 10^{-13} \msano $. General trends in mass-loss rate of  evolved low-mass stars provide loose constraints on mass-loss rates to be about $10^{-11}$--$10^{-9} \msano $  (see Figure \ref{fig:cranmer_land}), while Ly-$\alpha$ observations indicate a much lower mass-loss rate of $2 \times 10^{-15} \msano $ \citep{Wood2018}. Observed mass-loss rates are important for wind models, as they allow us to constrain the physical properties of wind. However, the large range of $\dot{M}$ derived so far in the literature has led to us to use a different approach to constrain the wind of $\lambda$~And, namely using radio observations.

Stellar winds can emit in radio through thermal bremsstrahlung emission. Our {polytropic wind} model predicts the largest thermal radio flux densities amongst all our simulations (0.89 mJy at 4.5 GHz). We have shown our Alfven-wave driven wind implementation struggled to reach the same level of radio emission, reaching at most 0.03 mJy. This is due to the fast exponential density decay in the lower atmosphere of our wave driven wind simulations. By comparing our radio calculations with radio observations we can constrain not only the mass-loss rate of \obj , but also the wind driving mechanism. There are two ways of doing that, depending on whether the observed radio emission is thermal or non-thermal. 

Considering that the quiescent observed radio emission of \obj\ is thermal, we can conclude that the wind of \obj\ is thermally driven with a mass loss rate of $3 \times 10^{-9}\msano$. This  is in line with mass-loss rates of neighbouring stars in the HR diagram (\Cref{fig:cranmer_land}). 
However, the quiescent observations we reported here does not rule out that this emission could be gyrosynchroton. In this case, we  turn our attention to the non-thermal flaring radio emission of \obj. If we are able to observe radio flares, then the wind of \obj\ must be optically thin, since radio flares were not absorbed by the wind, being able to escape the system. In this case, we cannot favour any of our two wind models and our conclusions are less stringent --  if this star has a thermally-driven wind, our  models indicate that the mass-loss rate of \obj\ is $\lesssim 3\times 10^{-9}\msano$. If the wind acceleration is caused by the dissipation of Alfven waves, then our upper limit in the mass-loss rate of \obj\ is one order of magnitude lower.  
}

Another interesting point to consider is that the wind from $\lambda$ And is quite variable, with chromospheric outflows not occurring when the star is faint (more starspots), and appearing when the star is bright \citep{Baliunas1979b}. Given more magnetic field maps (more observations in the future), the prospect of investigating cyclic behaviour and any correlations with other stellar and wind parameters would be quite exciting, and  this is something that could be addressed in future work. The study of cyclic behaviour through ZDI maps  \citep{BoroSaikia2016, 2018A&A...620L..11B, Jeffers2017, Jeffers2018}, and its effects on the winds \citep{Nicholson2016, Finley2019} have been done previously, but not for an evolved solar-mass star. Currently, radio observations are sparse and not well resolved temporally, it would be interesting to examine the trends in radio emission from the stellar wind and the derived magnetic geometry from ZDI maps.

\section*{Acknowledgements}
DOF and AAV would like to acknowledge funding from the European Research Council (ERC) under the European Union's Horizon 2020 research and innovation programme (grant agreement No 817540, ASTROFLOW). WM is supported by NASA grant NNX16AL12G. The authors wish to acknowledge the SFI/HEA Irish Centre for High-End Computing (ICHEC) for the provision of computational facilities and support. This work used the BATS-R-US tools developed at the University of Michigan Center for Space Environment Modeling and made available through the NASA Community Coordinated Modeling Center. The authors would like to thank {S. R. Cranmer} for access to stellar data used in  \Cref{fig:cranmer_land}. {The National Radio Astronomy Observatory is a facility of the National Science Foundation operated under cooperative agreement by Associated Universities, Inc. We thank the anonymous referees for their constructive feedback, which helped enhance the quality of our manuscript.}

\section*{Data Availability}
The data described in this article will be shared on reasonable request to the corresponding author.



\bibliographystyle{mnras}



\appendix

\section{ZDI adjustment of Stokes I and V LSD pseudo-profiles}
\begin{figure}
    \includegraphics[width=0.5\textwidth]{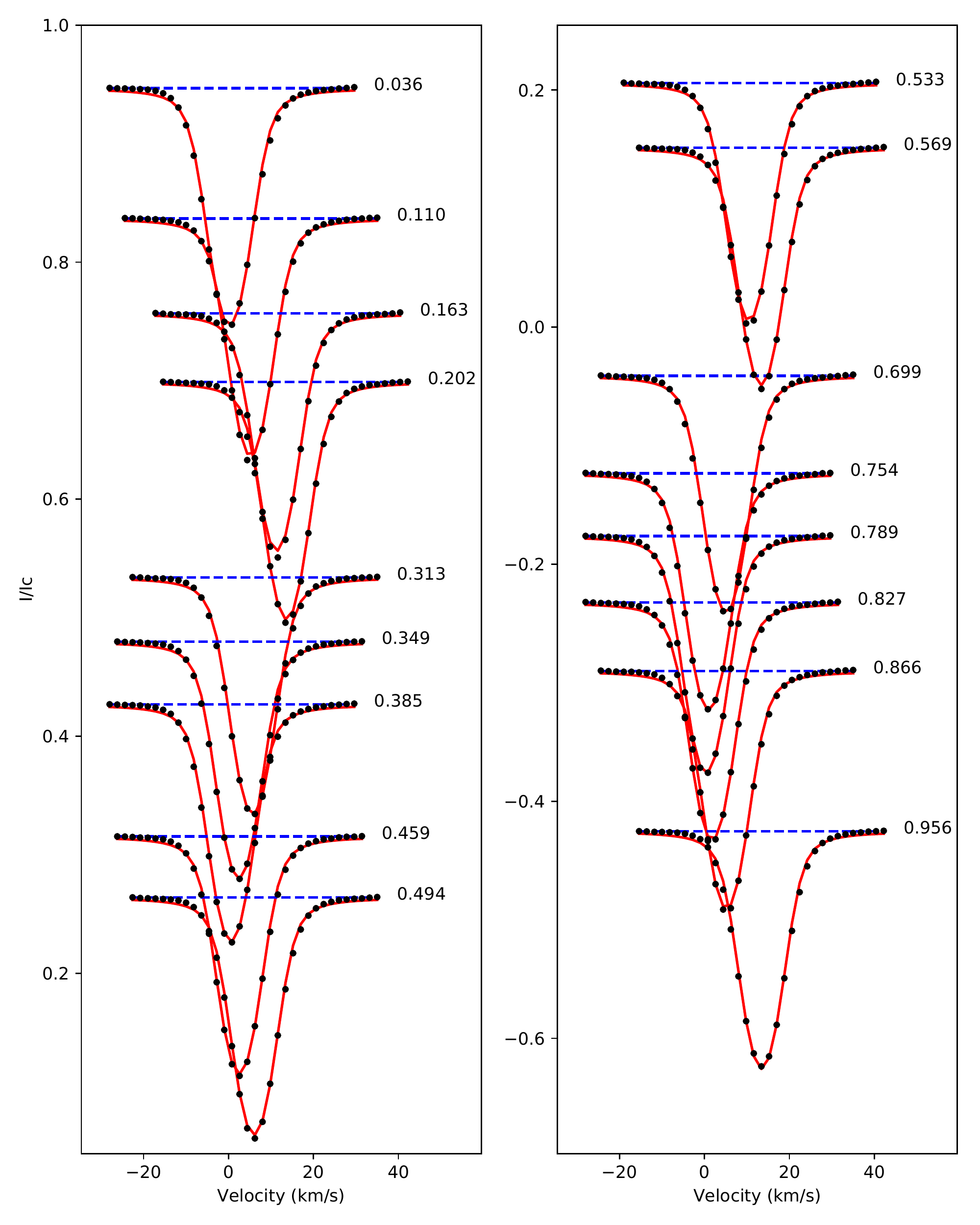}
  \caption{Stokes I LSD pseudo-profiles of $\lambda$~And (black points), over-plotted with the set of synthetic Stokes I profiles produced by the ZDI model (red line). The profiles are vertically shifted for display clarity. The dashed blue lines show the continuum level, and the phases of observation are indicated on the right of every profile, assuming a 54 d rotation period, and taking our first observation as phase origin. }
        \label{fig_stokesI}
\end{figure}

\begin{figure}
    \includegraphics[width=0.375\textwidth]{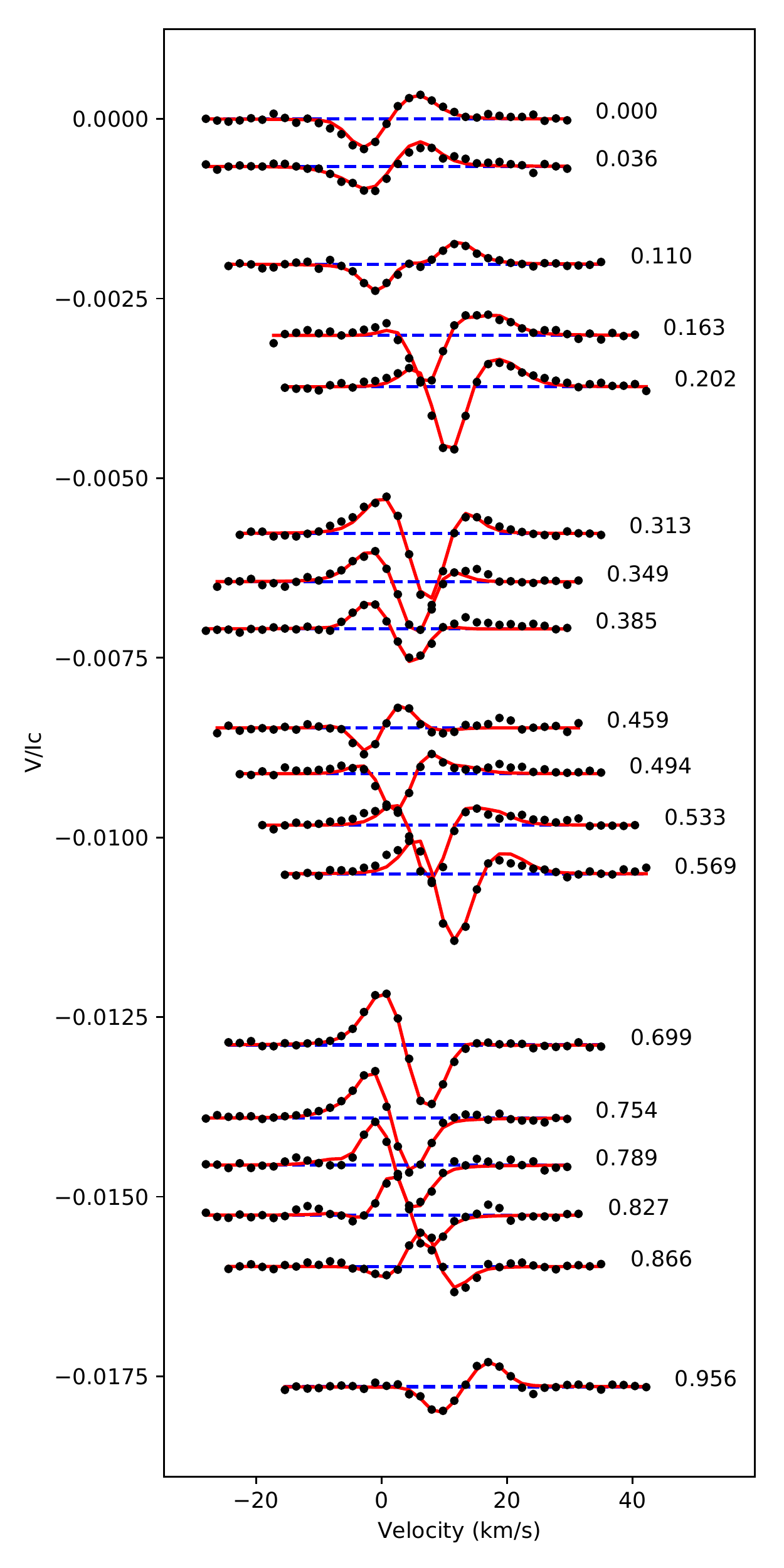}
  \caption{The same as Figure \ref{fig_stokesI}, but for Stokes V profiles.}        
  \label{fig_stokesV}
\end{figure}

{We  use  the LSD method \citep{Donati1997, 2010A&A...524A...5K} to extract an average, pseudo line profile of enhanced signal-to-noise ratio, which we then use to calculate the ZDI map presented in Section 2. We adopt a list of lines produced by a photospheric model \citep{Kurucz1993} with stellar parameters close to those of $\lambda$~And ($T_{\rm eff} = 4800 \pm 100$~K and $\log g = 2.75 \pm 0.25$, \citealt{Drake2011}). We impose for the LSD pseudo-line profiles an equivalent wavelength of 650~nm, and an equivalent Land\'e factor of 1.21. 
Figure \ref{fig_stokesI} shows the observed Stokes I LSD pseudo-profiles of $\lambda$~And (black points) over-plotted to the synthetic profiles produced with ZDI (red lines), while the Stokes V profiles are shown in Figure \ref{fig_stokesV}.}

\bsp	
\label{lastpage}
\end{document}